\documentclass[structabstract]{aa}
\usepackage{txfonts}
\usepackage{pictex}
\usepackage{latexsym}
\usepackage{graphics}
\usepackage{afterpage}

\begin{document}


\title
{
J004457+4123 (Sharov\,21): 
not a remarkable nova in M\,31 but a 
background quasar with a spectacular UV flare 
}

\author{H. Meusinger      \inst{1}
	 \and 
	 M. Henze    \inst{2}
         \and 
	 K. Birkle   \inst{3}$^{,}$\inst{4}
	 \and 
	 W. Pietsch  \inst{2}
	 \and 
	 B. Williams \inst{5}
	 \and 
	 D. Hatzidimitriou \inst{6}$^{,}$\inst{7}
         \and 
	 R. Nesci    \inst{8}
         \and 
	 H. Mandel   \inst{4} 
         \and\\ 
	 S. Ertel    \inst{9}
	 \and 
	 A. Hinze    \inst{10}
         \and 
	 T. Berthold \inst{11} 
	 }
	 
    \institute{
	 Th\"uringer Landessternwarte Tautenburg, Sternwarte 5, D--07778
         Tautenburg, Germany,
	       e-mail: meus@tls-tautenburg.de
         \and 
         Max-Planck-Institut f\"ur extraterrestrische Physik, Giessenbachstra\ss e,
         D--85748, Garching, Germany
         \and
	 Max-Planck-Institut f\"ur Astronomie, K\"onigstuhl 17, D--69117 
	 Heidelberg, Germany 
	 \and
	 ZAH, Landessternwarte Heidelberg, K\"onigstuhl 12, 
	 Universit\"at Heidelberg, D--69117 Heidelberg, Germany
	 \and
	 Department of Astronomy, Box 351\,580, University of Washington,
	 Seattle, WA 98195, USA 
	 \and
	 Department of Astrophysics, Astronomy and Mechanics, Faculty of 
         Physics, University of Athens, Panepistimiopolis, GR15784 Zografos, 
         Athens, Greece
	 \and
	 IESL, Foundation for Research and Technology, GR71110 Heraklion, 
	 Crete, Greece
         \and
     	 Department of Physics, University of Roma La Sapienza, Rome, Italy
	 \and
	 Institut f\"ur Theor.\,Physik und Astrophysik,
	 Christian-Albrechts-Universit\"at zu Kiel, 
	 Leibnizstra\ss{}e 15, D--24118 Kiel, Germany
	 \and
	 Astronomisches Institut,
	 Universit\"at Bern, 
	 Sidlerstra\ss{}e 5, CH--3012 Bern, Switzerland
	 \and
	 Sternwarte Sonneberg,
	 Sternwartestr. 32,
	 D--96515 Sonneberg, Germany
        }
	
\date{Received / Accepted }

\abstract {}
   {
   We announce the discovery of a quasar behind the disk of M\,31, which 
   was previously classified as a remarkable nova in our neighbour galaxy. 
   It is shown here to be a quasar with a single strong flare where the 
   UV flux has increased by a factor of $\sim 20$. The present paper is 
   primarily aimed at the remarkable outburst of J004457+4123 (Sharov\,21),
   with the first part focussed on the optical spectroscopy and
   the improvement in the photometric database. 
   }
   {
   We exploited the archives of photographic plates and 
   CCD observations from 15 wide-field telescopes and performed targetted new
   observations. In the second part, we try to 
   fit the flare by models of (1) gravitational microlensing due 
   to a star in M\,31 and (2) a tidal disruption event (TDE) of a star close 
   to the supermassive black hole of the quasar.}  
   {Both the optical spectrum and the broad band spectral energy distribution 
   of Sharov\,21 are shown to be very similar to that of normal, radio-quiet 
   type\,1 quasars.
   We present photometric data covering more than a century and resulting in a
   long-term light curve that is densely sampled over the past five decades.
   The variability of the quasar is characterized by a ground state with 
   typical fluctuation amplitudes of $\sim 0.2$\,mag around 
   $\bar{B} \sim 20.5$, superimposed by a singular flare of $\sim 2$\,yr duration 
   (observer frame) with the maximum at 1992.81. The total energy in the flare is 
   at least three orders of magnitudes higher than the radiated energy of the 
   most luminous supernovae, provided that it comes from an intrinsic process and 
   the energy is radiated isotropically.  The profile of the flare
   light curve is asymmetric showing in particular a sudden increase before the maximum,
   whereas the decreasing part can be roughly approximated by a $t^{-5/3}$ power law.
   Both properties appear to support the standard TDE scenario where a $\sim 10\,M_\odot$ 
   giant star was shredded in the tidal field of a $\sim 2\ldots5\,10^8\,M_\odot$ black hole. 
   The short fallback time derived from the observed light curve requires 
   an ultra-close encounter where the pericentre of the stellar orbit is deep
   within the tidal disruption radius.
   This simple model neglects, however, the influence of 
   the massive accretion disk, as well as general-relativistic effects on the orbit 
   of the tidal debris.       
   Gravitational microlensing probably provides an alternative 
   explanation, although the probability of such a high amplification event is very low.
   }
   {}
      
\keywords{Quasars: general --
          Quasars: individual: J004457+4123 --
	  Galaxies: individual: M\,31 --
	  Gravitational lensing --
	  Black hole physics
         }

\titlerunning{A spectacular UV quasar flare}
\authorrunning{H. Meusinger et al.}

\maketitle

%
%
\section{Introduction}
%
%

Temporal variability is one of the most conspicuous properties
for several classes of interesting astrophysical objects.
Owing to the unprecedented combination of sky coverage and photometric 
accuracy, discoveries from the Large Synoptic Survey Telescope (LSST),
the Panoramic Survey Telescope \& Rapid Response System (Pan-STARRS),
the Palomar-QUEST (PQ) survey, or the Palomar Transient Factory (PTF)
will provide great advances in the understanding of variable processes,
especially of rare transient phenomena 
(Gezari et al. \cite{Gezari08}; 
Strubbe \& Quataert \cite{Strubbe09};
Quimby et al. \cite{Quimby09}).
However, given the nature of the problem, the creation of the 
observational database for investigating the variability on 
long time scales in the reference frame of the source, as is the case
for quasars, takes a long time. This also holds for recurrent events, 
e.g. novae, with intrinsically shorter time scales where it is necessary 
to cover also the wide gaps between the single events. Presently, data 
mining in archives, most notably in the plate archives from large 
Schmidt telescopes, remains the only approach if the light curves have to 
cover a time interval of decades in the rest frame, at least for 
high-redshift quasars.

In the context of the search for and the identification of optically variable 
star-like sources in the field of the bright Local Group spiral galaxy M\,31 
(Pietsch et al. \cite{Pietsch05a}; Henze et al. \cite{Henze08}), our 
interest was pointed toward the apparent nova J004457+4123, originally 
discovered by Nedialkov et al. (\cite{Nedialkov96}) and described in more
detail by Sharov et al. (\cite{Sharov98}). The light curve presented by 
these authors clearly shows a strong bump with a maximum brightening by more 
than 3\,mag in the year 1992 while the source remained constant both in the 
23 years before and the 5 years after. Sharov et al. suggest that it is a 
``remarkable nova'' in M\,31, but underscore that it 
``differs dramatically from typical representatives of this class of objects''.  
Following the terminology of these authors (nova 21), we denote 
the object as Sharov\,21 throughout this paper. 
A possible X-ray counterpart was first discussed by Pietsch et al.
(\cite{Pietsch05b}), who searched for supersoft X-ray 
counterparts of optical novae in M\,31 and identified Sharov\,21 with the source 
[PFH2005]~601 of their catalogue of XMM-Newton EPIC X-ray sources 
(Pietsch et al. \cite{Pietsch05a}) and with the hard 
ROSAT source [SHL2001]~306 from the catalogue of Supper et al. (\cite{Supper01}).
Based on the hardness of the X-ray source and the peculiar optical light curve, 
Pietsch et al. (\cite{Pietsch05b}) speculate that Sharov\,21 
``may not be a nova at all''.  Here we present, for the first time, optical 
follow-up spectroscopy which reveals Sharov\,21 to be a quasar.

From the very beginning of the investigation of active galactic nuclei (AGN), 
variability is known to be a 
diagnostic property of this object class and has been successfully used  
as a criterion for the selection of quasar candidates in a number of studies 
(e.g., 
Kron \& Chiu \cite{Kron81};
Majewski et al. \cite{Majewski91};
Hawkins \& V\'eron \cite{Hawkins93};
Meusinger et al. \cite{Meusinger02}, \cite{Meusinger03};
Rengstorf et al. \cite{Rengstorf04}). 
AGN originally misclassified as variable stars are neither unprecedented nor 
unexpected. 
The most famous case is the prototypical blazar BL Lac,
discovered by Cuno Hoffmeister in 1930. 
However, the misclassification of a luminous quasar
as a nova is highly remarkable because it indicates 
a singular, strong outburst which points toward  
a rare and interesting transient phenomenon.

It has long been understood that the observed flux variations of AGNs 
hold keys to the structure of the radiation source.  
The physical mechanisms behind these fluctuations are however still 
poorly understood. Frequently discussed scenarios for the origin 
of the observed optical/UV broad-band long-term  (non-blazar) variability
related to massive or supermassive black holes in galaxy centres 
include various processes such as 
instabilities and non-linear oscillations of the accretion disk
(Taam  \& Lin \cite{Taam84}; 
Abramovicz et al. \cite{Abramovicz89}; 
Honma et al. \cite{Honma91};
Kawaguchi et al. \cite{Kawaguchi98}), 
multiple supernovae in the starburst environment
(Terlevich et al. \cite{Terlevich92};
Cid Fernandes et al. \cite{CidFernandes97}),
microlensing of the accretion disk or the broad line region 
by compact foreground objects 
(Chang \& Refsdal, \cite{Chang79};
Irwin et al. \cite{Irwin89};
Hawkins \cite{Hawkins93};
Schneider \cite{Schneider93};
Lewis \& Irwin \cite{Lewis96};
Zackrisson et al. \cite{Zackrisson03}),
the disruption of a star which passes within the tidal radius 
of the supermassive black hole 
(Hills \cite{Hills75}; 
Rees \cite{Rees88}, \cite{Rees90}; 
Komossa \& Bade \cite{Komossa99}; 
Komossa \& Meritt \cite{Komossa08}; 
Gezari et al. \cite{Gezari08}),
star-star collisions in the dense circumnuclear environment
(Torricelli-Ciamponi et al. \cite{Torricelli00}),
and interactions of the components in a supermassive binary black hole 
(Sillanp\"a\"a et al. \cite{Sillanpaa88};
Lehto \& Valtonen \cite{Lehto96};
Katz \cite{Katz97}; 
Liu \& Chen \cite{Liu07}).

The present paper is aimed at the highly peculiar light curve of the 
quasar Sharov\,21 which is worth detailed investigation. 
We present the optical spectrum and a significantly improved 
light curve and discuss possible scenarios for the strong outburst.  
The observations are described in Sect.\,\ref{obs}. The spectrum and other 
basic properties are analysed in Sect.\,\ref{prop}. 
The outburst is the subject of Sect.\,\ref{flare}.  
Two models are discussed in detail: gravitational microlensing and
a stellar tidal disruption event; alternative scenarios are briefly
summarized as well. Section\,\ref{conclusion} gives the conclusions. 
Standard cosmological parameters 
$H_0=71$ km s$^{-1}$ Mpc$^{-1}, \Omega_{\rm m}=0.27, \Omega_{\Lambda}=0.73$
are used throughout the paper.

%
%
\section{Observational data}\label{obs}
%
%

%
\subsection{Spectroscopy}
%

The optical spectrum was obtained with the Double Imaging Spectrograph (DIS)
on the 3.5-m telescope at Apache Point Observatory (APO) in New Mexico, USA
during a campaign to follow-up X-ray sources in M\,31.
Two exposures were taken for a total of 4\,500 seconds.
For the blue spectral range (3\,200 to 5\,500 \AA), the B\,400 reflectance 
grating was used with a dispersion of 1.85 \AA\ per pixel yielding a nominal 
resolution of about 7 \AA\ in combination with a 1\farcs5 entrance slit. 
The R\,300 grating, with a dispersion of 2.26 \AA\ per pixel, gives a resolution 
of 8 \AA\ for the red part (5\,000 to 10\,100 \AA).  The spectra were 
taken at UT 0300 on 2007-11-09. The observing conditions were excellent 
through the night.

The spectra were reduced, wavelength calibrated, and flux calibrated using 
the standard IRAF routines (ccdproc, identify, sunsfunc, apextract, apall).
Wavelength calibration was performed using HeNeAr lamp exposures taken just 
before the object exposures, and flux calibration was performed using a spectrum 
of the spectrophotometric standard star BD+28-4211.

%
\subsection{Optical photometry: long-term light curve}
%

The light curve published by Sharov et al. (\cite{Sharov98}) is based on 
B band observations taken with four telescopes between 1969.0 and 1997.7
with a good coverage of the outburst phase. The present study is aimed at an extended 
and better sampled long-term light curve. 
We exploited several data archives and combined the results with the data available 
in the literature. In addition, targetted new observations  for another 16 epochs in 
the years 2006 to 2009 were taken with the CCD Schmidt camera of the Tautenburg 2\,m telescope 
and with the focal reducer camera CAFOS at the 
2.2\,m telescope on Calar Alto\footnote{The Calar Alto Observatory of the
Centro Astron\'omico Hispano Alem\'an, Almer\'ia, Spain, is operated jointly by the
Max-Planck-Institut f\"ur Astronomie and the Instituto de Astrof\'isica de Andaluc\'ia
(CSIC).}, Spain. 
Most of the archival photographic plates were digitized in the frame of the present work 
using the Tautenburg Plate Scanner (Brunzendorf \& Meusinger, \cite{Brunzendorf99})
for the Tautenburg Schmidt plates, the high-quality commercial scanner at the Asiago 
observatory (Barbieri et al. \cite{Barbieri03}) for the Asiago plates, and the Microtek 
ScanMaker 9800XL for the Sonneberg astrograph plates.
The Calar Alto plates were scanned for the Heidelberg Digitized Astronomical 
Plates (HDAP) project using a Heidelberger Druckmaschinen Nexscan F4100 professional 
scanner and are available from the German Astrophysical Virtual Observatory 
(GAVO)\footnote{http://dc.zah.uni-heidelberg.de}.

\begin{table}[hhh]
\caption{Observational material for the construction of the light curve.}
\begin{flushleft}
\begin{tabular}{lrrlc}
\hline
telescope         &  $N_{\rm t}$    & $N_{\rm e}$   & years          & source\\ 
\hline
&&&&\\
\multicolumn{5}{l}{(a) Digitized photographic plates:} \\
Asiago Schmidt     &   24           &  8            & 1968.8--1993.1 & (1)\\
Calar Alto Schmidt &   43           & 15            & 1983.0--2000.7 & (1)\\
Calar Alto 1.2\,m  &    8           &  5            & 1976.7--1982.6 & (1)\\
Palomar Schmidt    &    4           &  4            & 1948.7--1989.7 & (1),(3)\\		   
Sonneberg 40\,cm   &    9           &  1            & 1992.2--1992.6 & (1)\\
Tautenburg Schmidt &  362           & 77            & 1961.5--1997.0 & (1)\\
&&&&\\
\multicolumn{5}{l}{(b) CCD observations:} \\
Calar Alto 2.2\,m  &    3           &  2            & 2008.7--2009.3 & (1)\\
CFHT 3.6\,m        &    1           &  1            & 1993.8         & (1)\\
INT (WFS)          &    5           &  1            & 1998.8         & (1)\\
INT 		           &  522           &  6            & 1999.7--2003.7 & (2)\\
Kitt Peak 4\,m     &   10           &  2            & 2000.8,2001.7  & (4)\\ 		    		
Skinakas 60\,cm    &    5           &  1            & 2007.6         & (1)\\
Tautenburg Schmidt &   30           & 14            & 2006.1--2009.7 & (1)\\
&&&&\\
\multicolumn{5}{l}{(c) Original data from Sharov et al. (\cite{Sharov98}):} \\
four other telescopes & $\sim 150$     & 84            & 1969.0--1997.7 & (5)\\  		
\hline
\end{tabular}\\
\end{flushleft}
\label{tab:all_observations}
\end{table}

A summary of all used observations from the last six decades is given in 
Table\,\ref{tab:all_observations} 
(CFHT = Canada France Hawaii Telescope, 
INT = Isaac Newton Telescope,
WFS = wide-field survey).
$N_{\rm t}$ is the total number of all single 
exposures, $N_{\rm e}$ the number of epochs in the light curve where the quasar 
has been measured. The last column gives the source of the photometric reduction:
(1) this work,
(2) Vilardell et al. (\cite{Vilardell06}), 
(3) Monet al al.  (\cite{Monet03}),
(4) Massey et al. (\cite{Massey06}),
(5) Sharov et al. (\cite{Sharov98}).

Altogether, the light curve data pool contains more than 1\,100
single observations from 15 telescopes.
Included are the B magnitudes published by Sharov et al. (\cite{Sharov98}) 
for 84 epochs, by the Local Group Galaxies Survey (LGGS; Massey et al. \cite{Massey06}) 
for 2 epochs, and by Vilardell et al. (\cite{Vilardell06}) binned
here into 6 epochs.  
For the other observations, the photometric reduction 
was done in the frame of the present study.
We used the Source Extractor 
package (Bertin \cite{Bertin96}) for object selection, background correction, 
and relative photometry, and the LGGS catalogue for the photometric calibration. 
The reduction was performed under ESO MIDAS.  
Because of the strongly inhomogeneous background across the disk of M\,31 
(Henze et al. \cite{Henze08}), the photometric calibration was done locally 
on a $8'\times 8'$ subimage around Sharov\,21 where typically $\sim 100 \pm 50$ 
calibration stars from the LGGS were identified. Note that 
the magnitudes given by Sharov et al. (\cite{Sharov98}) have also been derived 
from standard stars close to the target.
The blue magnitude for the Palomar POSS\,1 plate (1953-10-09) is taken from
the USNO-B1.0 catalogue (Monet et al. \cite{Monet03}) whereas the magnitudes 
given there for the POSS\,2 plate do not agree with the impression from the 
visual inspection of the images. We re-reduced the image cutouts from the 
Digitized Sky Survey (DSS2) and derived $B = 20.73\pm0.23$ and $R = 19.66\pm0.25$.
An early deep plate taken with the 1.2 m Samuel Oschin Telescope on 1948-09-29 
is shown in the {\it Hubble Atlas of Galaxies} (Sandage \cite{Sandage61}). The visual
inspection shows that Sharov\,21 is detected with $B = 20.3\pm0.3$. 

With very few exceptions the observations were made through filters reproducing 
the Johnson UBVR system. 
About 80\% of the data points in the light curve are from 
observations in the B band. 
From several pairs of observations, taken at nearly the same epoch 
but with different filters, we derive relations between 
colour indices and the B band magnitude. The results (Fig.\,\ref{fig:colour_indices}) 
are used to obtain $B$ from observations in the other bands.  
Because quasar variability is usually not achromatic (see Sect.\,\ref{prop}),
such colour relations are more suitable than single-epoch colour indices.
Photographic magnitudes $m_{\rm pg}$ from blue sensitive emulsions without filter 
were transformed by $B \sim m_{\rm pg}+0.1$.

There are two basic limitations with regard to the final set of data.
First, with $B>20$ for most of the time, Sharov\,21 is too faint to be 
detected in every plate archive. A high fraction of observations yields therefore upper limits only.
In other cases the detections are close to the plate limit resulting in relatively 
large photometric errors. Second, the epochs of the observations are 
not regularly distributed but show strong clustering. We stacked the 
images taken with the same telescope within typically several days
to a few weeks applying a quality-weighting factor (Froebrich \& Meusinger \cite{Froebrich00})
to obtain deeper images with improved signal-to-noise ratio. 
This procedure was not applied however for the outburst phase in 1992 where we are interested 
in a high temporal resolution.
 
\begin{figure*}[bhtp]   
\includegraphics{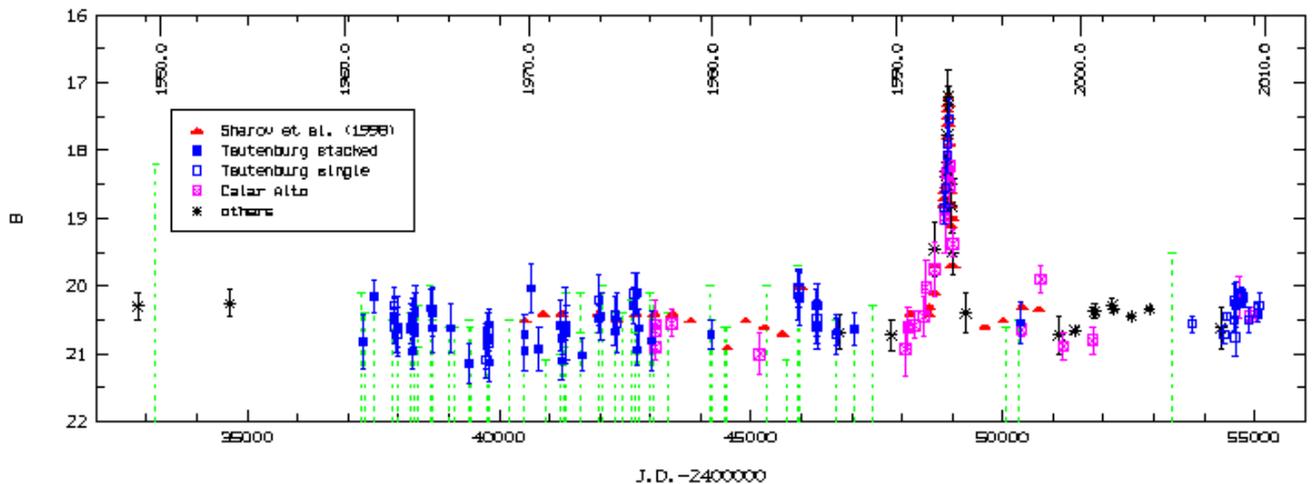}
\vspace{7.0cm}
\caption{
Long-term B light curve from the data summarized in Table\,\ref{tab:all_observations} 
(symbols plus error bars; no photometric errors available for the data from Sharov 
et al. (\cite{Sharov98})).
Dotted vertical lines: upper limits from images on which the object is not detected.
For lucidity, only a small fraction of the upper limit data is shown.
}
\label{fig:light-curve}
\end{figure*}

The final light curve (Fig.\,\ref{fig:light-curve}) comprises magnitudes at 221 detection 
epochs but still suffers from several gaps. In particular, no data are available 
for the early rising phase of the outburst between March and August 1992. We checked
the Wide Field Plate Datebase\footnote{Both the WFPDB version available via VizieR, 
(Tsvetkov et al. \cite{Tsvetkov97}), and the updated version available via the search 
browser developed in Sofia (http://draco.skyarchive.org/search) were used.}
but found no entries for this time. 
Also the search in the plate archives of the Baldone Schmidt telescope 
(Alksnis et al. \cite{Alksnis98})
and of the 100/300 cm Schmidt telescope of the Kvistaberg Observatory
revealed no observations of M\,31 during that time.

\begin{table}[hhh]
\caption{Detection limits on plates taken before 1950.
}
\begin{tabular}{lllc}
\hline
telescope & plate  &  date      &  $m_{\rm pg, lim}$\\
          &        & yyyy-mm-dd & \\
\hline
Bruce astrograph LHK & 23a   & 1900-09-14 & 18.8\\
Bruce astrograph LHK & 265a  & 1901-08-18 & 18.0\\
Yerkes 24~inch reflector & ? & 1901-09-18 & 19.5\\
Bruce astrograph LHK & 649a  & 1903-01-15 & 18.0\\
Bruce astrograph LHK & 842a  & 1903-09-27 & 18.0\\
Bruce astrograph LHK & 1384a & 1905-12-26 & 18.0\\
Waltz reflector LHK & 198   & 1907-11-02 & 17.5\\
Waltz reflector LHK & 369   & 1908-08-20 & 18.3\\
Waltz reflector LHK & 603   & 1909-10-19 & 18.5\\
Waltz reflector LHK & 4484  & 1934-09-17 & 18.5\\
Bruce astrograph LHK & 7163a & 1949-09-20 & 18.2\\
\hline
&\\
\end{tabular}\\
\label{tab:historical_plates}
\end{table}

It is useful to check also older historical observations of the Sharov\,21 
field which are not deep enough to detect the quasar in its faint stage but 
would allow discovering a previous outburst. 
Table\,\ref{tab:historical_plates} lists those observations from the 
40/200\,cm Bruce double-astrograph and the 72 cm Waltz reflector of 
the Landessternwarte Heidelberg-K\"onigstuhl (LHK) with $m_{\rm pg, lim} \sim 18$. 
A reproduction of a plate taken in 1901 with the 24 inch reflector at Yerkes observatory
is shown by Hubble (\cite{Hubble29}); from the visual inspection we estimate 
a detection threshold $m \sim 19.5$. Each of these observations excludes the 
occurrence of a flare similar to that of 1992 for at least several tens
of days around their dates of exposure.

%
%
\section{General properties of Sharov\,21}\label{prop}
%
%

In Sect.\,\ref{sec-spectrum} below we demonstrate that the previous classification 
of Sharov\,21 as a remarkable nova in M\,31 has to be rejected. Our optical spectrum,
presented below, clearly reveals the source to be a quasar. The most important 
properties of this quasar are summarized in Table\,\ref{tab:properties}.     
$t_{1/2}$ is the time interval for the decline from maximum
flux to half the maximum. Remarks:
(1) position from LGGS,
(2) ground state/maximum,
(3) observer frame/quasar rest frame,
(4) based on \ion{C}{iv} line,
(5) mean value for the ground state.

\begin{table}[hhh]
\caption{Basic properties of Sharov\,21 (remarks: see text).
 }
\begin{tabular}{lll}
\hline
\multicolumn{2}{l}{Measured and derived quantities}       & Remark\\
\hline
RA (2000)                           & $00^{\rm h}44^{\rm m}57\fs94$ &(1)\\
Dec (2000)                          & $+41\degr23\arcmin43\farcs$9  &(1)\\
redshift $z$                        & 2.109               &\\
projected distance from M\,31 centre  & 26\arcmin           &\\
apparent magnitude $B$              & 20.5/17.2           &(2)\\
foreground dust reddening $E(B-V)$  & 0.2 mag             &\\        
absolute magnitude $M_{\rm B}$      & $-27.5/-30.7$       &(2)\\
date of the maximum (year/JD)       & 1992.81/2448918     &\\
$t_{1/2}$ decline (days)            & 15/5                &(3)\\
black hole mass $M_{\rm bh}$        & $5\,10^8\,M_\odot$  &(4)\\
$\log\,(L_{\rm bol}/$ erg\ s$^{-1}$)& 46.6               &(5)\\
Eddington ratio $L_{\rm bol}/L_{\rm edd}$ & 0.60          &(5)\\
\hline
&\\
\end{tabular}\\
\label{tab:properties}
\end{table}

%
\subsection{Optical light curve and general remarks on variability}\label{variability}
%

\begin{figure}[bhtp]   
\includegraphics{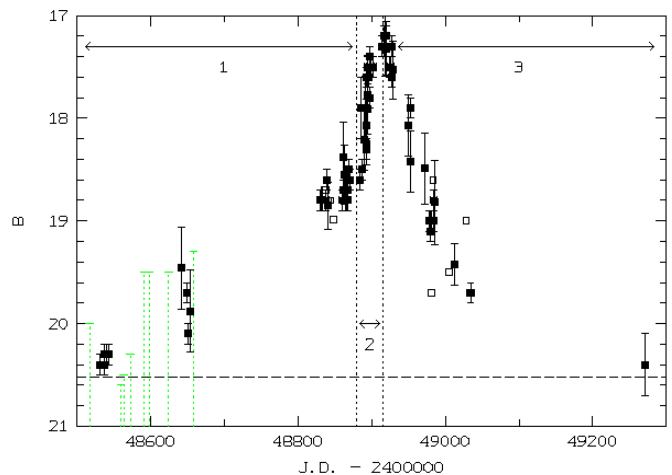}
\vspace{6.8cm}
\caption{ 
The light curve in the outburst. Open symbols: uncertain data, dashed lines: upper limits,
dotted horizontal line: ground state. 
} 
\label{fig:outburst}
\end{figure}

The light curve from the data discussed in Sect.\,\ref{obs} is shown in 
Fig.\,\ref{fig:light-curve}.
Sharov et al. (\cite{Sharov98}) note that the quasar was ``nearly constant from 1969
through 1991 with $B\approx 20.5$ and returned to this value one or two years after
the outburst''. Compared with the original data from Sharov et al., we
(1) basically confirm their finding, (2) extend the covered time 
interval by about one decade in each time direction, and (3) fill some broad gaps 
(e.g., between the years 1984 and 1990 and between the beginning
of 1993 and 1994). Based on the better sampling of our data, 
including the upper limits, the possibility of outbursts in intervals that 
were not covered by observations is thus significantly reduced. Hence, we conclude
that the light curve can be divided into (a) the faint state 
(with $\bar{B}$ = 20.52), 
which can be considered as the ground state, and (b) a single outburst, or flare, 
lasting $\sim 2$\,yr (JD $\sim 2\,448\,500 \ldots 2\,449\,300$) 
where the quasar was 3.3\,mag brighter in the maximum.
The flare (Fig.\,\ref{fig:outburst}) shows a slightly asymmetric 
profile with three phases: 
(1) a gradual increase between JD $\sim$2\,448\,500 and 2\,448\,880
with a gap in the light curve between March and August 1992, followed by 
(2) an abrupt rise to the maximum at  
JD $\sim$2\,448\,918, and (3) a quasi-exponential decline to the ground state 
at JD $\sim$2\,449\,300. The interpretation of the 
outburst will be the subject of Sect.\,\ref{flare}. 

To evaluate the variability in the ground state we compare Sharov\,21 
with quasars from the Tautenburg-Calar Alto Variability and Proper 
Motion Survey (VPMS; Meusinger et al. \cite{Meusinger02}, \cite{Meusinger03}). 
For Sharov\,21 we have a $B$ standard deviation
$\sigma_{\rm B} = 0.27$\,mag from the Tautenburg data
(0.26\,mag from all data). 
For the VPMS quasars with similar redshifts ($z = 2.1\pm 0.2$) and 
comparable (extinction-corrected) mean magnitudes
($\bar{B} = 19.7\pm0.2$) we have  $\sigma_{\rm B} = 0.26$ mag in the VPMS field around M\,3 
(8 quasars) and 0.29\,mag in the field around M\,92 (6 quasars). 
We conclude that the flux variability of Sharov\,21, in its ground state, 
is not unusually strong.

\begin{figure}[bhtp]   
\includegraphics{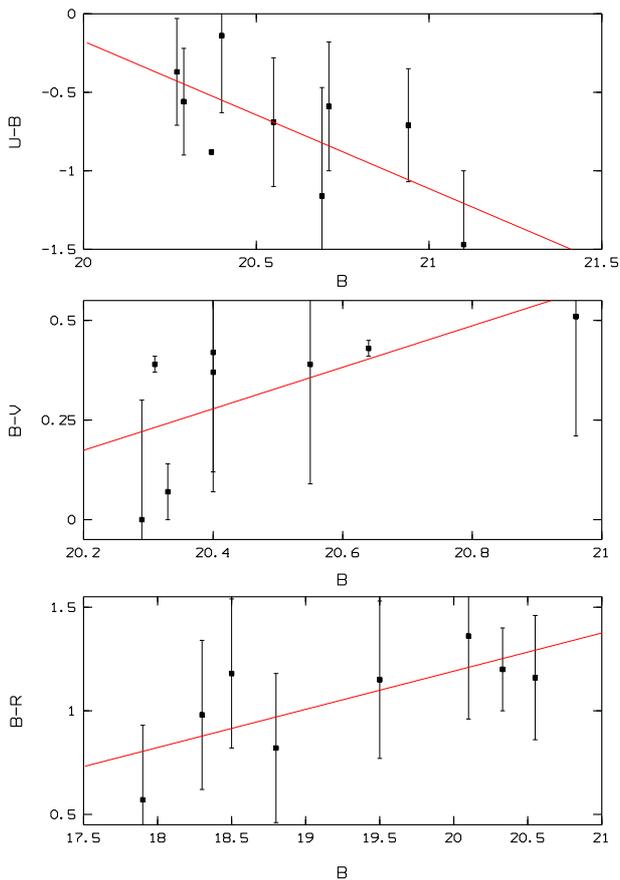}
\vspace{11.8cm}
\caption{
Colour indices of Sharov\,21 as a function of its B magnitude from
quasi-simultaneous observations in both bands.} 
\label{fig:colour_indices}
\end{figure}

Variations of the B band flux of Sharov\,21 are correlated with colour changes.
The observed relations (Fig.\,\ref{fig:colour_indices}) are qualitatively in 
agreement with the typical properties of quasars. A hardening of the optical/UV
continuum during the bright phase is indicated by multi-frequency monitoring of
selected AGNs 
(Cutri et al. \cite{Cutri85}; 
Edelson et al. \cite{Edelson90}; 
Paltani \& Courvoisier \cite{Paltani94})
as well as by statistical studies of AGN ensemble variability
(Di Clemente et al. \cite{DiClemente96}; 
Cristiani et al. \cite{Cristiani97}; 
Tr\`evese et al. \cite{Trevese01};
Vanden Berk et al. \cite{VandenBerk04}).
This trend has been confirmed also by multi-epoch spectroscopy of quasars from the Sloan 
Digital Sky Survey (SDSS) where it was shown that the emission lines are considerably less 
variable than the continuum, being stronger in the faint stage, relative to the continuum, 
than in the bright phase (Wilhite et al. \cite{Wilhite05}). 
For Sharov\,21, the U band is dominated by the 
strong Lyman\,$\alpha$/\ion{N}{v} line (Fig.\,\ref{fig:spectrum}). 
The contribution of the \ion{C}{iv} line to the
flux in the B band is much smaller, and the other bands are nearly 
pure continuum. The colour indices $B-V$ and $B-R$ are hence
expected to become bluer when the quasar becomes brighter, while 
$U-B$ becomes redder at the same time. Wilhite et al. present 
the colour differences between the bright and the faint phase as a 
function of redshift (their Fig.\,14)  indicating that $\Delta (u-g)$
has a local minimum at $z\approx 2$ while $\Delta (g-r)$ has a peak at the 
same redshift. Sesar et al. (\cite{Sesar07}) found a similar result from the 
multi-epoch photometric data of quasars in the SDSS stripe 82 with respect 
to the g and r bands. In an ongoing study of quasar variability in the SDSS 
stripe 82 (Meusinger et al., in preparation) we confirmed the relation shown 
by Sesar et al. and derived corresponding relations for the other bands which are 
in line with the results from Wilhite et al. (\cite{Wilhite05}).

%
\subsection{Position behind M\,31}
%

\begin{figure}[bhtp]   
\includegraphics{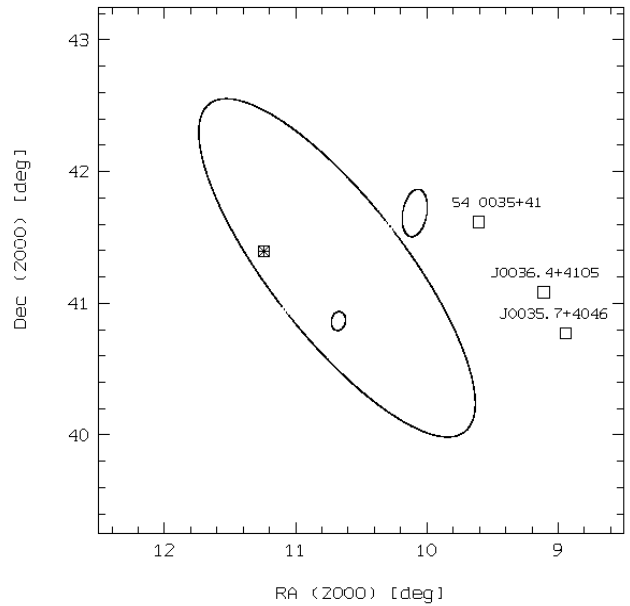}
\vspace{8.0cm}
\caption{
Map of the known quasars in the M\,31 field.
Open squares: quasars from V\'eron-Cetty \& V\'eron (\cite{Veron06}),
framed asterisk: Sharov\,21.
} 
\label{fig:position}
\end{figure}

Figure\,\ref{fig:position} reveals that Sharov\,21 is seen through the disk of M\,31. 
The ellipses represent the $D_{25}$ isophotes of M\,31, M\,32, and NGC\,205, 
respectively, according to the RC3 (de Vaucouleurs et al. \cite{deVaucouleurs91}). 
For  a distance of 750\,kpc for M\,31 (Vilardell et al \cite{Vilardell06}; their table\,1),
an inclination angle $i = 77\degr$, and a position angle of the major axis of $35\degr$ 
from the RC3, its projected distance from the centre of 26\arcmin\ (5.7\,kpc) corresponds 
to a galactocentric distance of $R = 16$ kpc in the midplane of M\,31.
A radius of $\sim 30$ kpc is a realistic assumption for the extent of the 
bright disk (e.g. 
Racine \cite{Racine91}; 
Ferguson et al. \cite{Ferguson02}; 
Irwin et al. \cite{Irwin05}). 
The ``Catalogue of Quasars and Active Galactic
Nuclei (12th Ed.)'' (V\'eron-Cetty \& V\'eron \cite{Veron06}) 
lists only three quasars within 100\arcmin\ from the centre of M\,31.
However, Sharov\,21 is, to our knowledge, the first quasar detected behind 
the disk of M\,31.

%
\subsection{Foreground reddening}
%

A foreground spiral galaxy is expected to produce a substantial 
reddening of a background quasar (\"Ostman et al. \cite{Ostman06}). 
The ``Galactic Dust Extinction 
Service''\footnote{http://irsa.ipac.caltech.edu/applocations/DUST/} 
of the NASA/IPAC Infrared Science Archive,
which is based on the method pioneered by Schlegel et al. (\cite{Schlegel98}), 
provides $E(B-V)=0.20$\,mag for the position of
Sharov\,21. Individual reddening values for a large set of  
globular clusters in M\,31 were derived by Barmby et al. (\cite{Barmby00}) 
and Fan et al. (\cite{Fan08}) yielding $E(B-V)=0.12\pm0.03$\,mag
for the six clusters within 6\arcmin\, from Sharov\,21. 
The reddening of the quasar must be stronger since only a fraction of the 
clusters is expected to be located behind the disk of M\,31. 
Finally, a simple model for the radial dependence of the extinction in M\,31
derived by Hatano et al (\cite{Hatano97}) yields $A_{\rm B} = 0.68$; i.e., 
$E(B-V)=0.17$ mag for the standard Milky Way extinction curve 
(Savage \& Mathis \cite{Savage79}), which seems to be valid also for M\,31 
(Barmby \cite{Barmby00}). Here we adopt $E(B-V)=0.20$\,mag for the total
foreground reddening of Sharov\,21.

%
\subsection{Optical spectrum}\label{sec-spectrum}
%

\begin{figure}[bhtp]   
\includegraphics{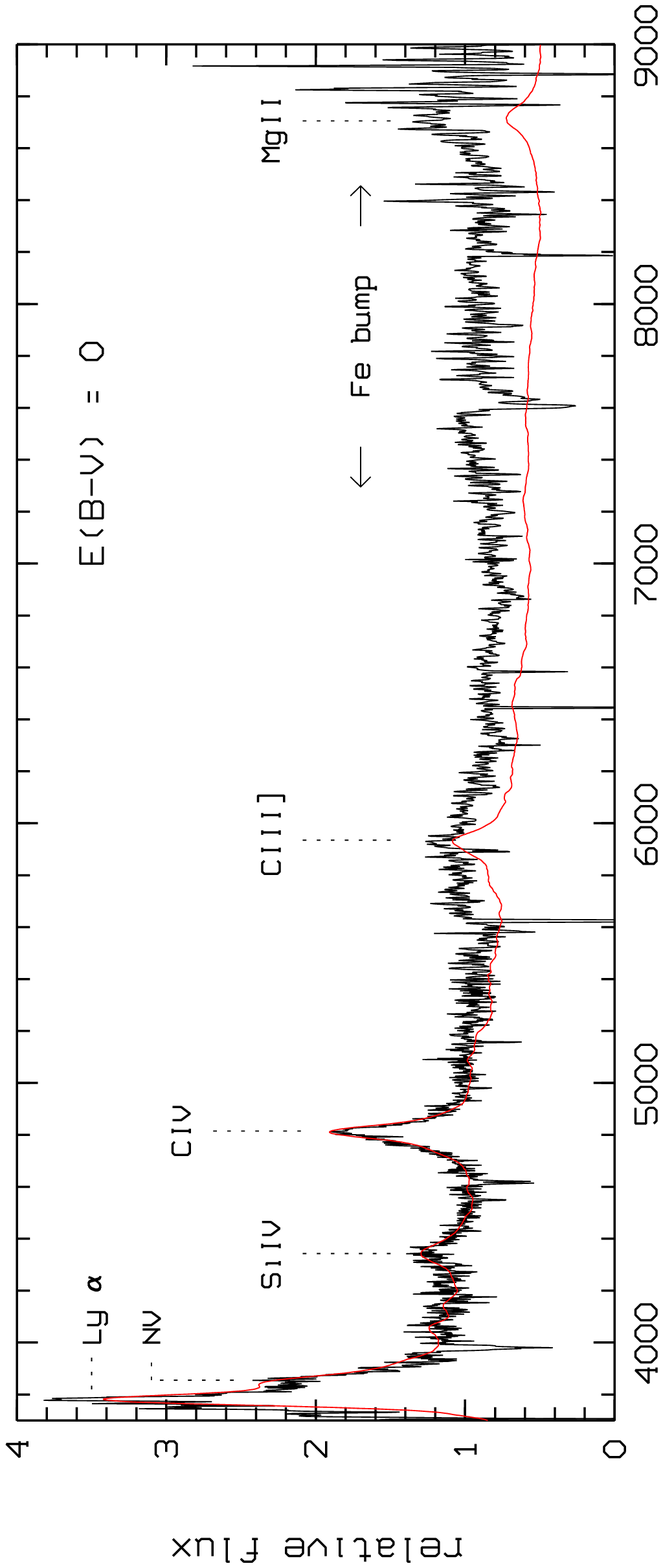}
\includegraphics{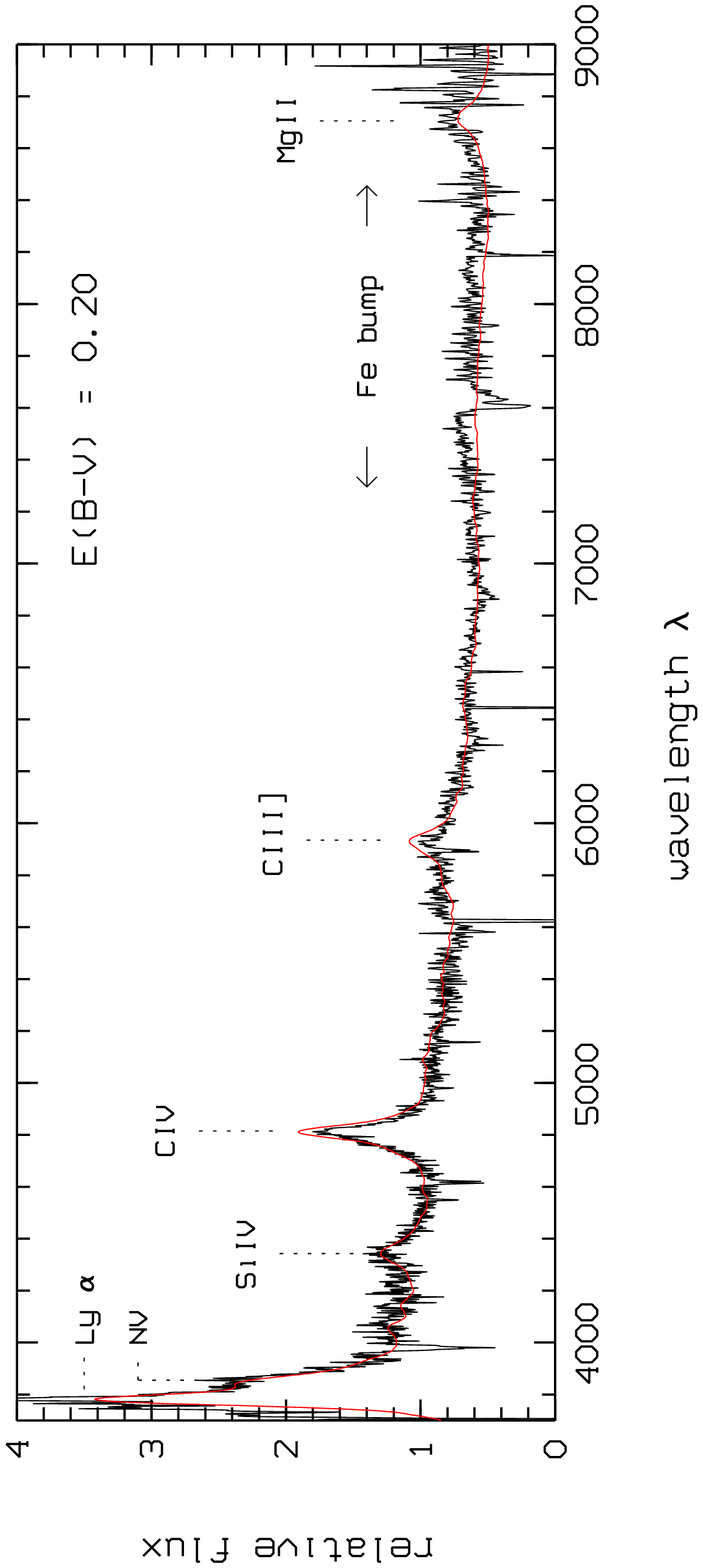}
\vspace{8.0cm}
\caption{Observed (top) and foreground extinction-corrected (bottom) optical spectrum 
of Sharov\,21 (observer frame), not corrected for telluric absorption. 
} 
\label{fig:spectrum}
\end{figure}

The optical spectrum is shown in Fig.\,\ref{fig:spectrum}.
For comparison the composite spectrum of ``normal'' quasars from the 
Sloan Digital Sky Survey (SDSS; Vanden Berk et al. \cite{VandenBerk01})
is plotted (thin smooth curve) shifted to the redshift of Sharov\,21,
the spectra are normalized at $\lambda\,4500$ \AA. 
We derived a redshift of $z=2.109$ both from the fit of the SDSS composite 
and directly from the wavelengths of the narrow components of the 
Lyman\,$\alpha$ and \ion{N}{v} lines. 
Compared with the SDSS composite Sharov\,21 has a redder continuum.
We de-reddened the spectrum for foreground ($z=0$) extinction adopting 
the Milky Way extinction curve. Good agreement with the mean SDSS 
quasar spectrum is found for $E(B-V) = 0.2$ mag (Fig.\,\ref{fig:spectrum}, 
bottom), which is perfectly in line with the reddening value from the 
NASA/IPAC Infrared Science Archive (see above). 

The de-reddened spectrum of Sharov\,21 is that of a typical type\,1 quasar.
There is no evidence of unusual spectral features indicating a peculiar 
nature of Sharov\,21. Compared with the SDSS composite the reddening 
corrected spectrum shows a stronger \ion{Fe}\ bump at 
$\lambda \sim 7300\ldots 8300$ \AA\ (observer frame) which points
towards a relatively high Eddington ratio (Dong et al. \cite{Dong09}).
The \ion{C}{iii]} $\lambda\,1909$ \AA\, line appears slightly weaker, 
but note that the line coincides with the \ion{Na}{i} foreground 
absorption at $\lambda\lambda\,5890,5896$ \AA. We notice further a weak 
unidentified absorption line at the position of the 
$\ion{N}{iv} \lambda\,1486$ \AA\ line.

%
\subsection{Other wavelength regimes}\label{multiwave}
%

M\,31 has been observed at radio wavelengths both as part of larger surveys and as the 
focus of dedicated programmes (see e.g., Gelfand et al. \cite{Gelfand04}; their Table 1). 
In no case, a radio counterpart was detected at the position of Sharov\,21.
With a flux density limit of 0.15 mJy, the VLA survey by Braun (\cite{Braun90}) 
is the deepest one at 20\,cm continuum (1.465\,GHz). At lower radio frequencies 
a deep survey of the M\,31 field was performed by 
Gelfand et al. (\cite{Gelfand04}, \cite{Gelfand05}) 
with the VLA yielding a flux density limit of 3\,mJy  at 325\,MHz. 
Within $d \sim 2$\arcmin\ from Sharov\,21, the 325\,MHz catalogue 
lists the sources GLG\,043, 045, 050, and 051. 
However, with $d>1$\arcmin\, all four radio sources are obviously 
not related to Sharov\,21.
The non-detection in the Braun survey implies an 
upper limit for the radio-loudness parameter $R$,
i.e. the ratio of the 5 GHz radio flux density to the 2500 \AA\ optical
flux density in the quasar rest frame (Stocke et al. \cite{Stocke92}),  
log\,$R^{\ast} < 0.7$ for a radio spectral index $\alpha_{\rm R} = -0.3$ and 
log\,$R^{\ast} < 0.9$ for $\alpha_{\rm R} = -0.5$. With the threshold
log\,$R^{\ast} \ge 1$ for radio-loud AGNs (e.g., White et al. \cite{White00})
Sharov\,21 is not a radio-loud flat-spectrum quasar. 

We re-analysed the archival XMM-Newton and ROSAT data. For computing fluxes from instrument 
dependent count rates, we used an absorbed power-law model with a generic photon index of 
1.7 (see also Pietsch et al. \cite{Pietsch05a}). We adopted the foreground 
extinction of $E(B-V)=0.20$\,mag, derived form the optical data, which 
translates to a \hbox{$N_{\rm H}$}~ of 1.1$\, 10^{21}$ cm$^{-2}$ 
following Predehl et al. (\cite{Predehl95}). Based on this spectral model, we 
used the source count rates given by Pietsch et al. (\cite{Pietsch05a}) for 
[PFH2005]~601 in the XMM-Newton observation 0151580401 (2003-02-06) to estimate 
an unabsorbed flux of ($5.8\pm1.9$) $\, 10^{-14}$ \hbox{erg s$^{-1}$ cm$^{-2}$} 
in the (0.2 - 10.0) keV band and the monochromatic flux $F_{\nu} = (4.5\pm1.5)\,10^{-9}$\,Jy
at $\nu = 1.7\,10^{17}$\,Hz.

The ROSAT data of the Sharov\,21 field\footnote{see  
http://www.xray.mpe.mpg.de/cgi-bin/rosat/seq-browser} consists of 12 observations with the 
Position Sensitive Proportional Counter (PSPC) and 3 observations with the High Resolution 
Imager (HRI). The data analysis was done under ESO MIDAS within the EXSAS context. We 
performed source detection around the position of Sharov\,21 on the original event files 
and computed count rates and 3$\sigma$ upper limits for all observations. There is just 
one 3$\sigma$ detection of an X-ray source in the data set, identical with [SHL2001]~306, 
which is supplemented by upper limits for the rest of the observations. The flux estimated 
from the detection is consistent, within the errors, with the XMM-Newton data. Although 
the ROSAT observations were performed around JD 2449000, i.e. during the decline of the UV 
flare, no significant X-ray variability is detected. Note however, that the only ROSAT 
detection is a very faint off-axis PSPC detection and due to the large positional error 
circle of this instrument we can not assume a doubtless correlation with [PFH2005]~601.

\begin{figure}[bhtp]   
\includegraphics{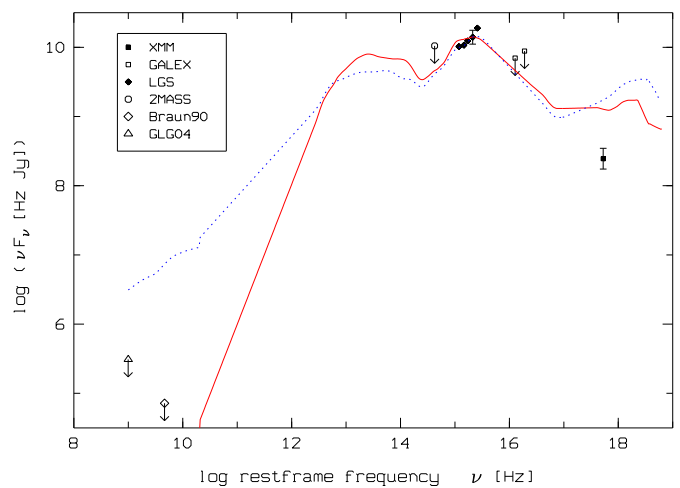}
\vspace{6.8cm}
\caption{ 
SED of Sharov\,21 in the restframe (open symbols with downward arrows: upper limits). 
For comparison the mean SED (Elvis et al. \cite{Elvis94}) is shown
for radio-quiet (solid) and radio-loud quasars (dotted), 
normalized at $\lambda$\,1415 \AA.  
} 
\label{fig:sed}
\end{figure}

We also checked images from the Deep Imaging Survey with the Galaxy Evolution Explorer 
(GALEX) in the near ultraviolet ($\lambda_{\rm eff} \sim 2270$ \AA) and in the far 
ultraviolet  ($\lambda_{\rm eff} \sim 1520$ \AA), but no counterpart could be 
identified within a radius of $\sim$ 5\,\arcsec. For detections in the near infrared we searched 
in the 2MASS catalogue (Skrutskie et al. \cite{Skrutskie06}), again without a
clear-cut identification of a counterpart. From simple
statistics of the sources around the position of Sharov\,21 a flux limit of
1 $\mu$Jy is estimated for the ultraviolet bands and of 0.25 mJy in the K band.
In Fig.\,\ref{fig:sed} we compare the extinction-corrected 
broad band spectral energy distribution (SED)
of Sharov\,21 with the mean SED for normal, nonblazar quasars from Elvis et al.
(\cite{Elvis94}).

%
\subsection{Black hole mass and Eddington ratio}
%

The black hole mass is estimated from the \ion{C}{iv} line width as a measure 
of proxy for the velocity dispersion of the broad emission line gas in combination 
with a radius-luminosity ($R$-$L$) relationship for the emission region. 
Significant progress has been achieved over the last years with the calibration 
of the $R$-$L$ relation 
(Vestergaard \cite{Vestergaard02}, \cite{Vestergaard09}; 
Corbett et al. \cite{Corbett03};
Warner et al. \cite{Warner03};
Peterson et al. \cite{Peterson04};
Vestergaard \& Peterson \cite{Vestergaard06}).

We use equation (8) from Vestergaard \& Peterson (\cite{Vestergaard06})
which is based on the line dispersion $\sigma_{\rm line}(\mbox{\ion{C}{iv}})$ 
and the monochromatic continuum luminosity $\lambda L_\lambda$ at 1\,350\,\AA.
Following the method outlined by Peterson et al. (\cite{Peterson04}) and
Vestergaard \& Peterson (\cite{Vestergaard06}) we obtain 
$\sigma_{\rm line}(\mbox{\ion{C}{iv}}) = (2.9\pm0.5)\,10^3$ km s$^{-1}$. 
The luminosity of the continuum at $\lambda = 1\,350$\,\AA\ (restframe) is derived from
the extinction-corrected mean B band flux in the ground state after 
correction for the contribution from the emission lines ($\sim 12$\%) 
and assuming a standard power-law continuum  
$F_\lambda \propto \lambda^{-(2+\alpha)}$ with $\alpha = -0.5$.
We find $\lambda L_\lambda (1\,350\,\mbox{\AA}) = 1.07\,10^{46}$ erg\ s$^{-1}$ 
which finally yields a black hole mass of $M_{\rm bh} \sim 5\,10^8\,M_\odot$. 
Vestergaard \& Peterson (\cite{Vestergaard06}) give
a standard deviation of $\pm0.33$ dex for the scaling relation. 
Allowing further for the uncertainties of
the line width and of the continuum luminosity, the total uncertainty 
of $M_{\rm bh}$ is roughly a factor of 3. 

Adopting the bolometric correction $k_{\rm bc}(1\,350\,\mbox{\AA}) = 4$ 
from Richards et al. (\cite{Richards06}; their Fig.\,12), the 
monochromatic luminosity from above corresponds to 
$
L_{\rm bol} = \lambda L_\lambda \cdot k_{\rm bc} 
            = 4.3\,10^{\,46}$ erg\,s$^{-1} 
	     \sim 10^{13} L_\odot
$ 
and to an Eddington ratio 
$
\epsilon \equiv L_{\rm bol}/L_{\rm edd} 
             = 3.3\,10^{8} \ (M_{\rm bh}/M_\odot)^{-1} 
	     \sim 0.6 
$
for the ground state. 
The overwhelming majority of SDSS quasars with $z\sim 2$ 
radiate at between 10\% and 100\% of their Eddington luminosity 
(see Vestergaard \cite{Vestergaard09}) with a narrow
$\epsilon$ interval at high luminosities (Kollmeier et al. \cite{Kollmeier06}; 
Shen et al. \cite{Shen08}; Onken \& Kollmeier \cite{Onken08}).
According to Onken \& Kollmeier, quasars with
$\log\,(L_{\rm bol}/\mbox{erg\ s}^{-1}) >46.5$ and $z = 1.1\ldots2.2$ 
are characterized by $\langle \log\,\epsilon \rangle = -0.65\pm0.35$ whereas 
$\epsilon$ is smaller and shows a broader distribution for low-luminosity AGNs.  
Gavignaud et al. (\cite{Gavignaud08}) find
$\log\,\epsilon = -0.97+0.28\,[\log\,(L_{\rm bol}/\mbox{erg\ s}^{-1})-45]$ 
over the interval $\log\,(L_{\rm bol}/L_\odot) = 45\ldots48.5$. 
As can be seen from their Fig.\,4, a value of 
$\epsilon \sim 0.6$ is relatively high but not unusual for 
quasars of the redshift and luminosity of Sharov\,21. However, 
in the maximum of the outburst, where the B band flux is a factor of 
$\sim 20$ higher, the luminosity of Sharov\,21 corresponds to a highly 
super-Eddington regime.

It has been questioned for a couple of reasons whether the mass estimates 
from scaling relations over-predict the black hole masses by about 
an order of magnitude. For arguments in favour of the mass estimates we
refer to the discussion by Vestergaard (\cite{Vestergaard04}). In addition 
we note that a significantly lower black hole mass would require an 
unusually high Eddington ratio for Sharov\,21.

%
%
\section{The outburst}\label{flare}
%
%

\subsection{Sharov\,21-like UV flares are rare}\label{Rareness}

UV Flux variations by one or a few magnitudes are observed both in low-luminosity 
AGNs like NGC\,5548 (Ulrich et al. \cite{Ulrich97}, and references therein) 
or in blazars (Sect.\,\ref{other_interpretations}). 
Contrary to the burst of Sharov\,21, the B band flux of high redshift 
radio-quiet quasars typically varies by a few tenths of a magnitude 
as is observed in the ground state of Sharov\,21. 
The second remarkable difference is the fact that the strong activity of
our quasar is limited to a short time interval of $\sim 1$ year (observer frame). 
During the remaining $\sim 47$ yr covered by the light curve (i.e., the ground state), 
the mean flux variation is a factor of $\sim 16$ smaller than the 
maximum fluctuation in the outburst. In contrast, both strongly variable 
low-luminosity AGNs like NGC\,5548 and optically violently variable (OVV) 
quasars show a more or less steady up and down variation.

\begin{figure}[bhtp]   
\includegraphics{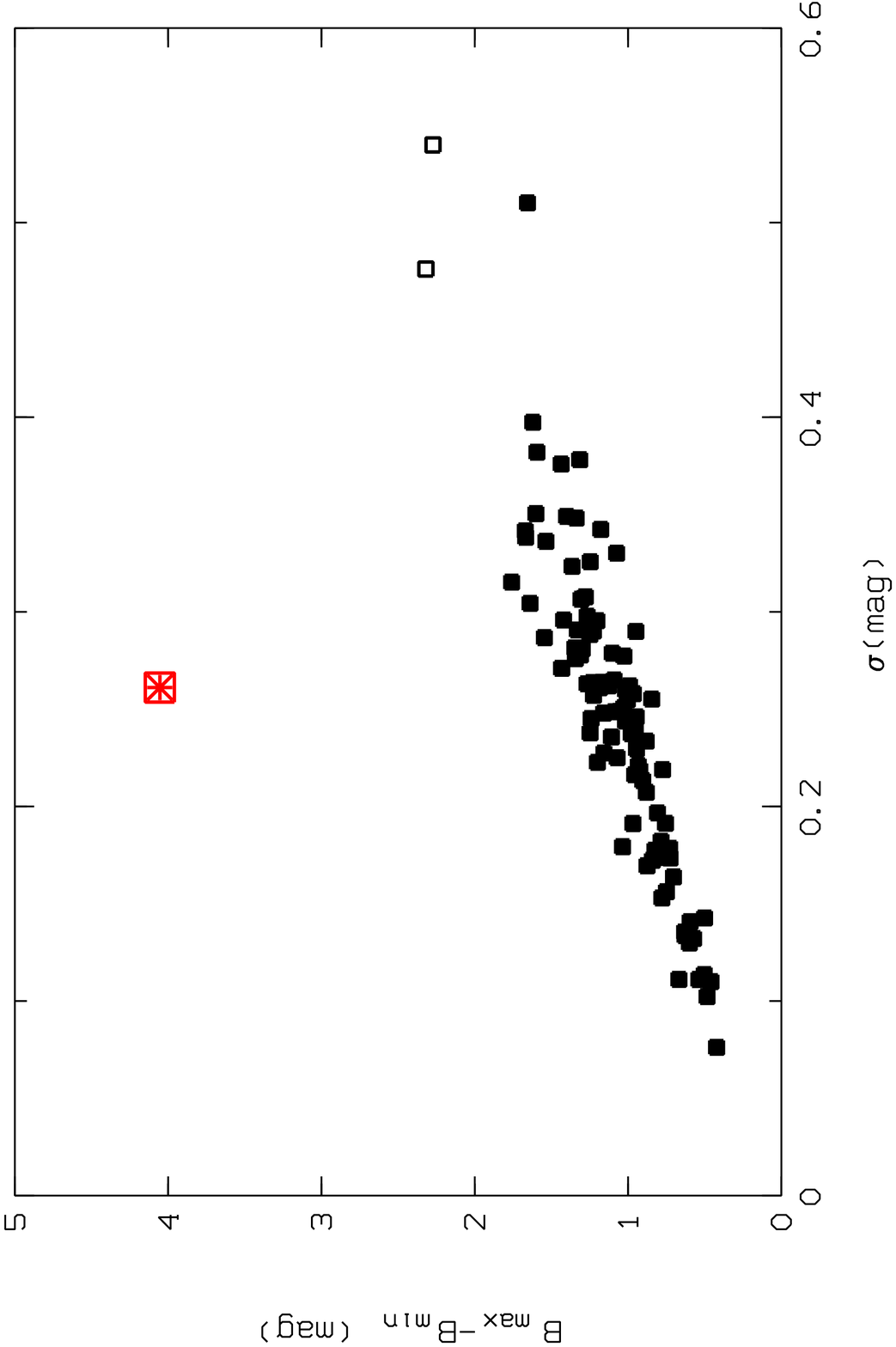}
\includegraphics{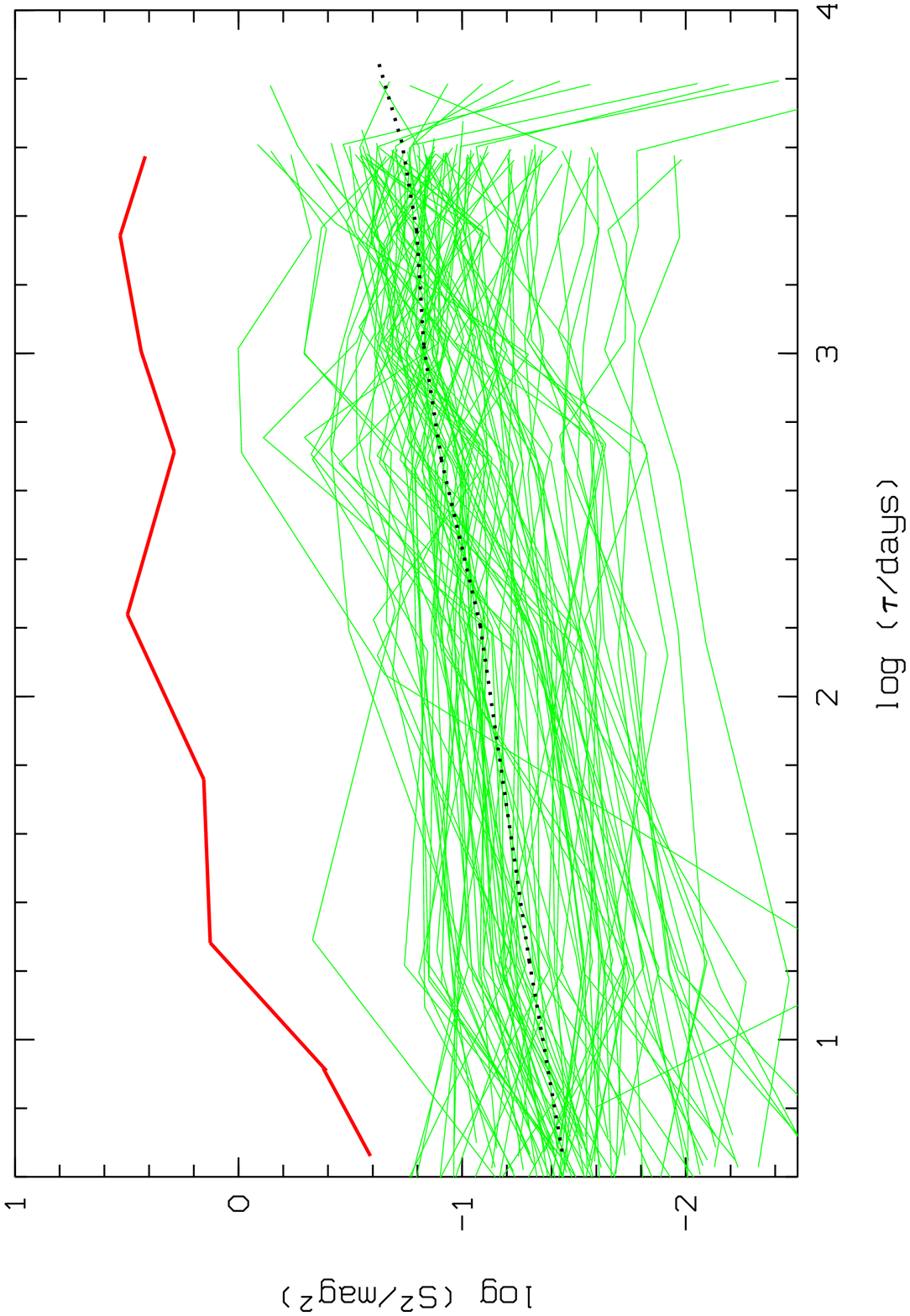}
\vspace{12.5cm}
\caption{
Comparison of the variability of Sharov\,21 with the VPMS quasars.
Top: maximum fluctuation $\Delta B_{\rm max}$ {\it vs.} standard 
deviation $\sigma$ (framed asterisk: Sharov\,21; $\sigma$ refers to the ground state).
Bottom: single object structure functions for Sharov\,21 (solid line at the top) and the 
VPMS quasars (thin solid lines), and VPMS sample-averaged structure function 
(dotted line).
} 
\label{fig:sf_vpms}
\end{figure}

To get an idea how usual or unusual the outburst of Sharov\,21 is, 
we first consider the light curves of the quasars from the Variability and 
Proper Motion Survey (VPMS; 
Meusinger et al. \cite{Meusinger02}, \cite{Meusinger03}). 
The time baseline of the observations is nearly the same 
(from $\sim 1960$ to 2008) as for Sharov\,21, the numbers of observations 
per quasar are however much lower, typically $\sim 50$. The VPMS 
quasar sample is highly complete down to $B\sim 20$. Among the 321 AGNs
in the VPMS there are 10 AGNs with maximum fluctuations 
$\Delta B_{\rm max} = B_{\rm max}-B_{\rm min} \ge 2$ mag. For the majority 
of them (80\%), the variability
is characterized by a more or less monotonic variation over the time 
baseline. Only for two AGNs  the maximum fluctuations 
can be attributed to burst-like features. One is the strongly variable
object CC\,Boo, a Seyfert galaxy at $z=0.17$ (Margon \& Deutsch \cite{Margon97}), 
the other one is a radio-quiet quasar at $z = 1.08$. 
However, with $\Delta B_{\rm max} = 2.46$ mag and 2.50\,mag, respectively, 
the burst amplitudes are considerably smaller than for Sharov\,21 and
there are strong fluctuations also in other parts of the light curves. 

In Fig.\,\ref{fig:sf_vpms} we compare variability properties of 
Sharov\,21 with those of the VPMS quasars. The top panel shows the 
maximum fluctuation amplitude versus standard deviation of the B magnitudes
in the ground state for Sharov\,21, compared with 98 VPMS quasars
of comparable redshifts ($z = 2.1\pm 0.5$). As we cannot reasonably 
distinguish between a ground state and a higher state for the majority
of the quasars we simply use the standard deviation of all data in the 
light curve as a proxy. (A natural consequence is the increase of $\sigma$
with the maximum amplitude.)  The light curves of the two strongly variably 
VPMS quasars marked by open squares in Fig.\,\ref{fig:sf_vpms} 
are clearly dominated by smooth long-term variations over decades.
A popular statistical tool for the investigation of quasar variability
is the first order structure function 
$S^2(\tau) = \langle [m(t+\tau)-m(t)]^2 \rangle_{\rm t}$
(e.g., Simonetti et al. \cite{Simonetti85};
Kawaguchi et al. \cite{Kawaguchi98})
where $\tau$ is the time-lag between two observations in the quasar
restframe and the angular brackets denote the time-average.
The structure function represents a sort of running 
variance (as a function of the time-lag) and contains therewith
information about the time scales of the involved variability 
processes. The most important conclusion from Fig.\,\ref{fig:sf_vpms} is
that the flare of Sharov\,21 is singular and without comparison
in the long-term variability data of the VPMS quasar sample. 

Excellent data for the statistical study of quasar variability has been provided
from stripe 82 of the Sloan Digital Sky Survey (SDSS) 
for $\sim10^4$ quasars in five colour bands over $\sim7$ years 
(e.g., Sesar et al. \cite{Sesar07}). 
Using the SDSS quasar catalogue (Schneider et al. \cite{Schneider07})
we identified 8311 quasars in the Light and Motion Curve Catalogue 
(LMCC; Bramich et al. \cite{Bramich08}) from $\sim\,249$ square degrees 
of the SDSS stripe 82. No high-redshift quasars with $z>2$ 
were found with amplitudes in the
u and g bands $> 1.5$ mag. Allowing for the whole redshift range,
the two SDSS quasars with the highest amplitudes in the g band are
SDSS\,J001130.0+005751.8 ($z = 1.49$) and 
SDSS\,J211817.37+001316.8  ($z = 0.46$)
with $\Delta g_{\rm max} =$ 3.2 and 2.7\,mag, respectively.
Both are bright polarized flat-spectrum radio sources 
(Jackson et al. \cite{Jackson07}; Sowards-Emmerd et al. \cite{Sowards05}) 
and their variability is hence characterized by blazar activity
(Sect.\,\ref{other_interpretations}). Interestingly, both 
show (1) a trend of reddening when they become brighter
and (2) a trend of increasing intrinsic variability 
(for definition see Sesar et al. \cite{Sesar07}) with increasing wavelength.
Such a behaviour is opposite to typical radio-quiet quasars and also 
to Sharov\,21 (Sect.\,\ref{variability}). 

Results from the Palomar-QUEST Survey were recently presented by 
Bauer et al. (\cite{Bauer09}). 3\,113 objects were identified in
7\,200 square degrees with fluctuation amplitudes $> 0.4$ mag on time scales up to
$\sim 3.5$ yr. There are only a few objects showing maximum amplitudes 
$> 2$ mag up to 3.7 mag; all of them are blazars. Nearly all of the 14\,800 
spectroscopically identified quasars in the data base 
have jumps $<1$ mag; the highest value is 1.8 mag.

\subsection{A microlensing event?}

As the flare of Sharov\,21 appears to be a singular feature in the
long light curve, it is tempting to speculate that it originates from 
a rare event. Here we first discuss microlensing.

Chang \& Refsdal (\cite{Chang79}) first suggested that the flux of a (macrolensed)
quasar can be affected by a star crossing close to the line of sight 
with time scales of the order of a few months to several years.
On the observational side, quasar microlensing was first identified 
by Irwin et al. (\cite{Irwin89}). Since then, 
considerable progress has been made in microlensing simulations, 
and observations of significant microlensing have been reported in a 
number of systems
(e.g., 
Pelt et al \cite{Pelt98}; 
Koopmans et al. \cite{Koopmans00};
Chae et al. \cite{Chae01};
Wisotzki et al. \cite{Wisotzki03}, \cite{Wisotzki04};
Chartas et al. \cite{Chartas04};
Eigenbrod et al. \cite{Eigenbrod06};
Paraficz \cite{Paraficz06};
Sluse et al. \cite{Sluse07}).

\begin{figure}[bhtp]   
\includegraphics{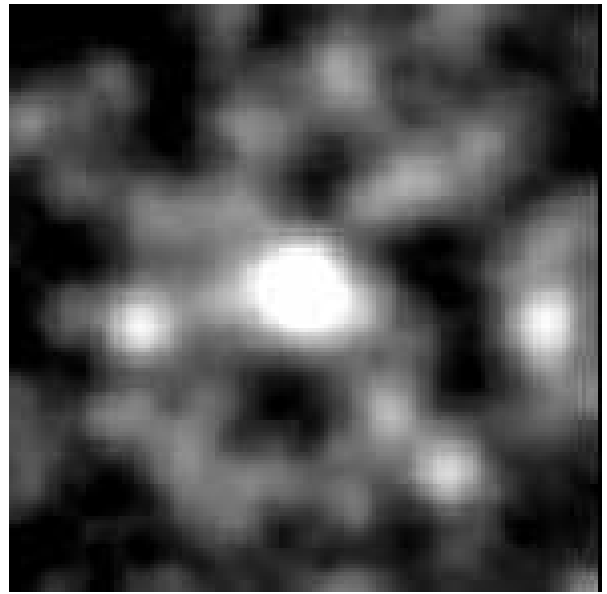}
\vspace{9.1cm}
\caption{
The crowded field towards Sharov\,21 on the 
10\arcsec $\times$ 10\arcsec cutout from the 4-m KPNO image
in the I band (N up, E left).
} 
\label{fig:neighbourhood}
\end{figure}

Sharov\,21 is seen through M\,31, the high star density close to the 
line of sight is illustrated by Fig.\,\ref{fig:neighbourhood}. 
The quasar is the brightest, slightly elongated object in the centre,
all other objects are most likely stars in M\,31. Note also that the 
quasar appears slightly elongated which points towards an object at a distance 
$\la 0$\,\farcs5. Unfortunately, there are no archival 
Hubble Space Telescope observations of the field. Extensive imaging was performed
with the WFPC2 in 2008 to create an accurate map of M\,31 microlensing 
(PI: A. Crotts) where Sharov\,21 is, however, several  
arcseconds out of the field.  A deep image of the field 
was taken with Subprime-Cam at the 8 m Subaru telescope in 2004, but 
the quasar lies exactly in the gap between the two adjacent fields 6 and 7. 

From the M\,31 mass model (Geehan et al. \cite{Geehan06})
we estimate a stellar column density of $\Sigma_\ast \sim 170\,M_\odot$ pc$^{-2}$
towards the quasar. This value is $\sim 2.5\,10^{-4}$ smaller than the critical 
surface density $\Sigma_{\rm crit} = D c^2/(4\pi G)$ with 
$D = D_{\rm S}/(D_{\rm L} D_{\rm LS})$ and $D_{\rm LS} = D_{\rm S}-D_{\rm L}$,
where $D_{\rm S}$ and $D_{\rm L}$ are the angular distances of the source and the lense, 
respectively, $G$ is the gravitational constant. 
The optical depth, i.e. the probability for the quasar to fall into the Einstein
radius of a star in M\,31, is  
$\tau = \Sigma_\ast/\Sigma_{\rm crit} \sim 2.5\,10^{-4}$.

In the case of quasar microlensing by stars in a foreground galaxy with
high optical depth the lenses do not act individually and the 
light curve is complex. For low optical depth ($\tau \la 0.5$), 
however, microlensing can be studied in the single-star approximation 
(Paczy\'nski \cite{Paczynski86}), which is applied here. More precisely, 
the assumptions are made that both the 
lens and the source are point-like and that the relative motion of the lens is
linear. Then the light curve is given by the magnification 
\begin{equation}\label{Eqn:magnification}
\mu(t) \equiv \frac{F_{\nu,\rm obs}(t)}{\bar{F}_{\nu,\rm gs}} 
= \frac{u(t)^2+2}{u(t)\sqrt{u(t)^2+4}},
\end{equation}
where $u(t)$ is the angular distance between source and lens in units of the 
Einstein angle $\Theta_{\rm E}$, and
$F_{\nu,\rm obs}(t)$ and $\bar{F}_{\nu,\rm gs}$ are the  
observed monochromatic flux density at time $t$ and the mean flux density in the 
ground state, respectively, in the B band.    
As $D_{\rm S}$ and $D_{\rm L}$ are given, the light curve
depends only on the mass $M_{\rm L}$, the relative transverse velocity 
$v_{\rm t}$ of the lens, and the impact parameter, i.e., the 
minimum distance $u_{\rm min}$ between lens and source. 
The light curve of a high magnification event can be significantly modified 
by the finite size of the source. This, however, occurs for 
$u_{\rm min} \sim R_\ast/R_{\rm E}$, with $R_{\rm E} = \Theta_{\rm E} D_{\rm L}$, 
whereas we have $u_{\rm min} \sim 10^3\,R_\ast/R_{\rm E}$. 
Furthermore, for a scale of 8.4\,kpc/\,\arcsec\ at the redshift of Sharov\,21 and
assuming that the source size of the UV radiation of the quasar
is $\la 10^{13}$ m,  the source has an angular diameter about two orders 
of magnitude smaller than the minimum impact parameter and can be considered
as point-like.

The transverse velocity is determined by the motion of the lens in M\,31,
the proper motion of M\,31 with respect to the barycentre of the Local Group (LG),
the motion of the Sun around the Galactic centre, the motion of the Galaxy 
around the Local Group barycentre, and the motion of the LG relative 
to the cosmic microwave background (CMB). Since the first two effects are poorly
constrained, we consider here for simplicity only the velocity of the 
LG with respect to the CMB. With $v_{\rm LG-CMB} = 612$ km\,s$^{-1}$ towards
$(l,b) = (270\degr,29\degr)$ (Loeb \& Narayan \cite{Loeb08}) we have
$v_{\rm t} \sim 300$ km\,s$^{-1}$.

\begin{figure}[bhtp]   
\includegraphics{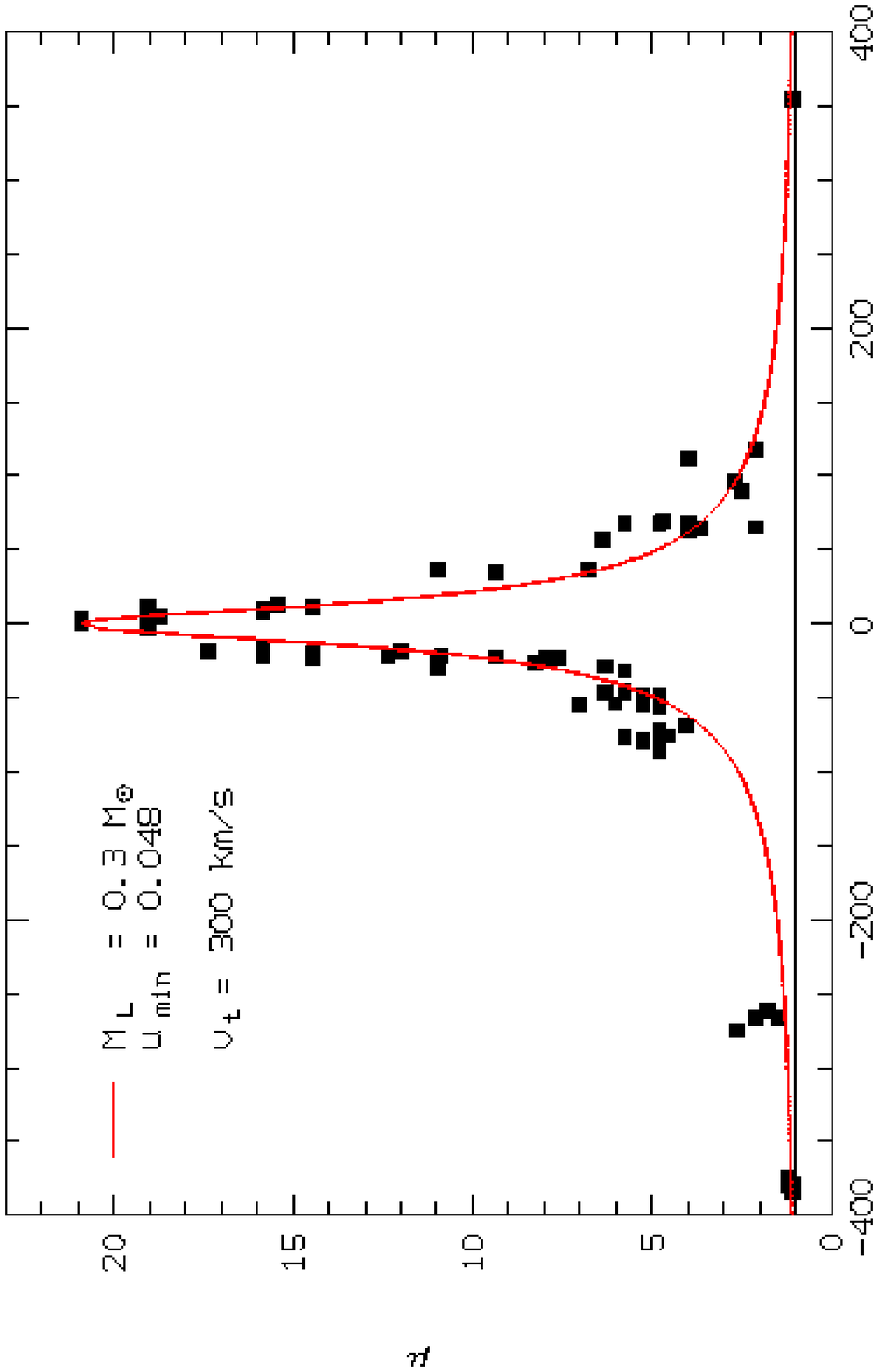}
\includegraphics{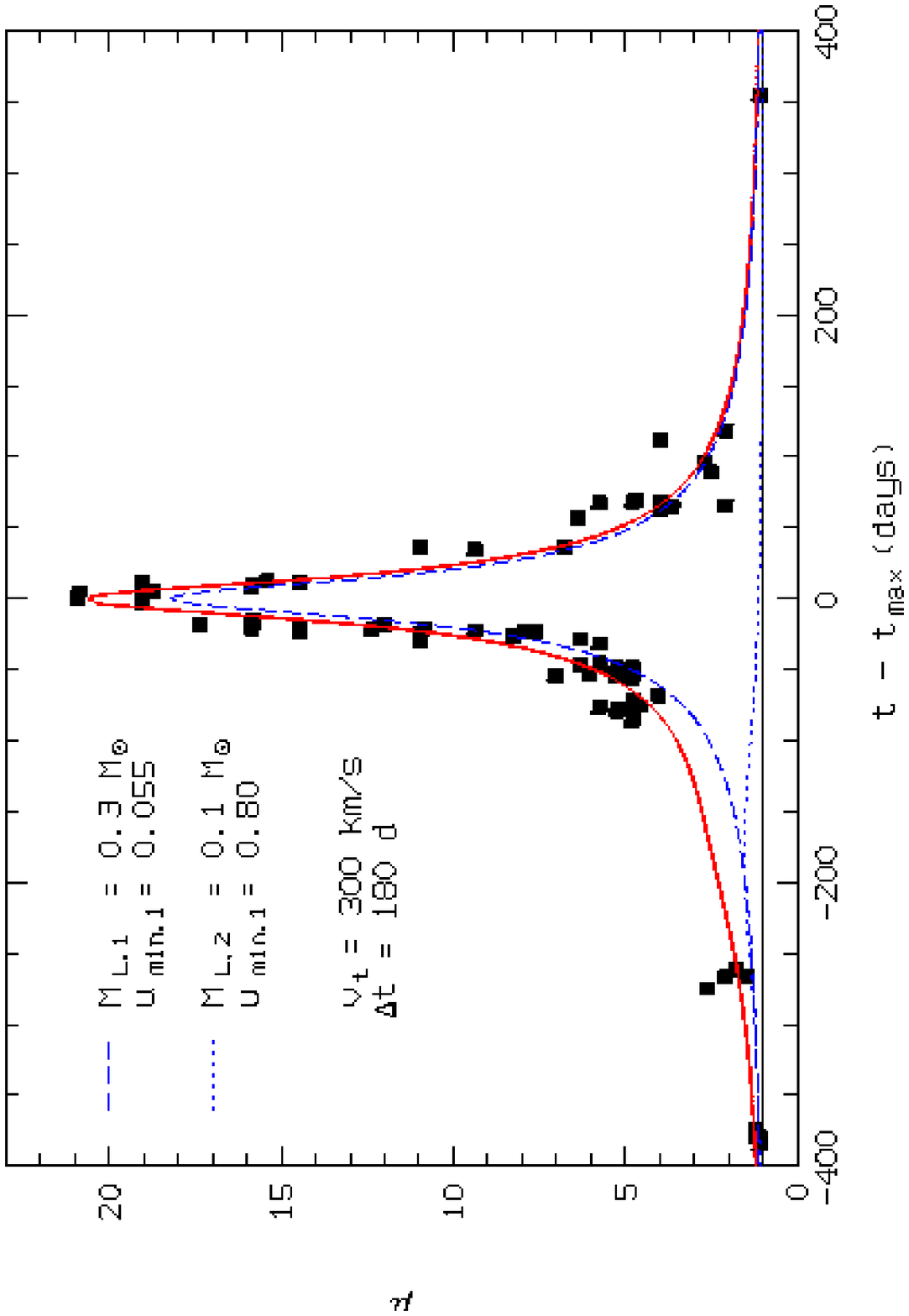}
\vspace{12.0cm}
\caption{Light curve $\mu(t)$ of Sharov\,21 around its maximum 
(no error bars for lucidity, open symbols: uncertain data). Overplotted are the
fitting curves (solid) for microlensing by one single star (top)
and two single stars of the masses $M_{\rm L,1}$ and $M_{\rm L,2}$ (bottom).  
} 
\label{fig:microlensing}
\end{figure}

For the simplifying assumption that all lenses have the same mass and velocity,
the Einstein radius crossing time $t_{\rm E} = \Theta_{\rm E}/v_{\rm t}$
is constant and the event rate for quasar microlensing is estimated by 
$\Gamma \sim 2\,N_{\rm q} \tau/(\pi t_{\rm E})$ (e.g., Mao \cite{Mao08}), 
where $N_{\rm q}$ is the number
of quasars. Given $\sim 20$ quasars with $B < 20$ per square degree
and a surface area of $\sim 2$ square degrees within the 25\,mag isophote
of M\,31, we have $N_{\rm q} \sim 40$ and $\Gamma \sim 1$ per century.
Hence, the discovery of a microlensed, faint quasar behind M\,31 over an interval 
of half a century is not unlikely.

When the tranverse velocity and the distances are fixed, 
the maximum amplification is determined by the impact parameter 
$u_{\rm min}$ 
and the time scale $t_{\rm E}$ is determined by $M_\ast$.  
For $v_{\rm t} = 300$ km\,s$^{-1}$ the 
light curve is best fitted with $M_\ast \sim 0.3\,M_\odot$ and $u_{\rm min} \sim 0.048$
(Fig.\,\ref{fig:microlensing}, top). A higher velocity requires a higher stellar 
mass. The fit is not perfect because the observed light curve shows
a weak but clearly indicated asymmetry (Fig.\,\ref{fig:outburst}) 
which hints on deviations from the single-lens hypothesis. 
Therefore we add a second lens which moves with the same $v_{\rm t}$ but 
crosses the line of sight $\sim 180$ days earlier. For simplicity, each component is 
treated as a single star, the light curve is computed via the multiplicative
magnification approximation (Vietri \& Ostriker \cite{Vietri83})
$\mu = \mu_1 \cdot \mu_2$.
A good fit to the shoulder in the light curve 
between 400 and 50 days before the maximum is achieved for 
$(M_{\ast,1}, M_{\ast,2}) = (0.3\,M_\odot, 0.1\,M_\odot)$ 
if $v_{\rm t} = 300$ km\,s$^{-1}$ and  
$(1.2\,M_\odot, 0.4\,M_\odot)$ 
if $v_{\rm t} = 600$ km\,s$^{-1}$, respectively, with
$(u_{\rm min,1}, u_{\rm min,2} = 0.055,0.8)$ in both cases 
(Fig.\,\ref{fig:microlensing}).   
The projected linear separation of the two lenses is $\sim 35$ AU. Both the 
mass ratio and the separation are not atypical of a binary star.

Although microlensing by stars in M\,31 appears to be a plausible explanation 
for the flare of Sharov\,21, there are serious objections. 
First, the probability for magnification as strong as in Sharov\,21 is 
very low. With regard to his model with $\tau = 0.1$,
Paczy\'nski (\cite{Paczynski86}) points out 
that ``appreciable increases in intensity are there once per millennium ...
there is not much hope in detecting intensity changes due to microlensing
at such low optical depth''. 
Second, colour index variations are expected in the case
of quasar microlensing (Wambsganss \& Paczy\'nski \cite{Wambsganss91};
Yonehara et al. \cite{Yonehara08}) if the source appears extended. 
For the point source-point lens constellation, however, an
achromatic light curve is expected. Here, the source 
can be considered point-like, but $B-R$ was considerably smaller in the 
flare compared to the faint state (Fig.\,\ref{fig:colour_indices}).
Third, as the angular separation of the two lenses is $\sim \Theta_{\rm E,1}$,
the magnification pattern is expected to be more complex then for the 
single-star approximation. Detailed modelling with a more consistent treatment 
of the binary lens problem is clearly necessary but is beyond the scope of 
the present paper. It will be particularly interesting to see if such models 
find a natural interpretation for the steep rise of the flux before the peak.

%
\subsection{A stellar tidal disruption event?}
%

An alternative process which is rare and produces a strong UV/X-ray flare 
is the disruption of a star by the strong tidal forces of a massive black hole 
(Lidskii \& Ozernoi \cite{Lidskii79}; Rees \cite{Rees88}, \cite{Rees90}).   
Tidal disruption events (TDEs) were discussed so far mainly in the context of 
dormant black holes in non-AGN galaxies or in low-luminosity AGNs 
(e.g., Phinney \cite{Phinney89}; Rees \cite{Rees90}) 
and, more recently, of recoiling black holes 
(Komossa \& Meritt \cite{Komossa08}). At least a fraction of low-luminosity AGNs 
appears to be powered by stellar tidal disruptions 
(Komossa et al. \cite{Komossa04}; 
Milosavljevi\'c et al. \cite{Miloslavljevic06}). For high-luminosity AGNs 
the situation is more complicated. TDEs are expected to be
rare for black hole masses higher than a critical mass 
$M_{\rm bh} > M_{\rm crit} \approx 10^8\,M_\odot$ where the  
gravitation (Schwarzschild) radius, $R_{\rm S}$, exceeds the tidal disruption 
radius of solar mass stars so that such stars are swallowed whole without disruption
(Hills \cite{Hills75}; see also Chen et al. \cite{Chen08}; 
Gezari et al. \cite{Gezari08}). On the other hand,
it seems possible that a massive, self-gravitating accretion disk
brings more stars into loss-cone orbits and enhances therefore the tidal 
disruption rate (Syer et al. \cite{Syer91}; Donley et al. \cite{Donley02}).

Until now, about a dozen TDE candidates in non-AGN galaxies have been found 
from X-ray surveys
(Komossa \cite{Komossa02};
Komossa et al. \cite{Komossa09}; 
Esquej \cite{Esquej07}, \cite{Esquej08}; 
Cappelluti et al. \cite{Cappelluti09}) and also in the UV/Optical
(Renzini et al. \cite{Renzini95};
Gezari et al. \cite{Gezari06}, \cite{Gezari08}).
None of the events detected so far were found to be related to a high-luminosity AGN. 
Such a detection would be interesting because it provides an opportunity
to check the basic tidal disruption theory as their predictions generally depend on 
$M_{\rm bh}$ which can be estimated independently in this case.  
However, as noted by Gezari et al. (\cite{Gezari08}), the existence of various 
mechanisms for the UV variability of quasars makes the interpretation of a UV 
flare subject to careful analysis. In particular it is necessary to make sure
that the observed flare is not just a more or less usual feature in a strongly
variable light curve. 

The theoretical frame for the interpretation of TDEs has been set with the pioneering work by 
Hills (\cite{Hills75}), 
Lacy et al. (\cite{Lacy82}),
Rees (\cite{Rees88}), 
Phinney et al. (\cite{Phinney89}), 
Evans \& Kochanek (\cite{Evans89}),
and more recently by, among others,
Magorrian \& Tremaine (\cite{Magorrian99}),
Ulmer (\cite{Ulmer99}),
Ayal et al. (\cite{Ayal00}),
Ivanov \& Novikov (\cite{Ivanov01}),
Menou \& Quataert (\cite{Menou01}),
Li et al. (\cite{Li02}),
Bogdanovi\'c et al. ({\cite{Bogdanovic04}),
Wang \& Merritt (\cite{Wang04}), 
Chen et al. (\cite{Chen08}),
Lodato et al. (\cite{Lodato08}).
We start with a short summary of the basic theory.
A star of mass $M_\star$ and radius $R_\star$, originally in hydrostatic 
equilibrium and on a parabolic orbit, passing at the time $t_{\rm d}$ 
within the tidal disruption radius 
\begin{equation}\label{Eqn:tidal_radius}
R_{\rm t} = R_\ast (M_{\rm bh}/M_\ast)^{1/3}
\end{equation} 
of the black hole will be torn apart by the tidal forces. The change in the
black hole potential over the star will produce a spread in the specific energy
of the gas. After the encounter, the gas particles will have a considerably widened
energy distribution. Following Rees (\cite{Rees88}), about half of the mass has 
negative energy and is thus gravitationally bound to the black hole, the other half 
is unbound and will be ejected
from the system. The bound elements move in highly excentric elliptical orbits of the size
$2a = R_{\rm p} + R_{\rm a} \sim R_{\rm a}$ where $R_{\rm p}$ and $R_{\rm a}$ 
are the pericentric and the apocentric distance, respectively. With 
the Keplerian relation between the energy and the size of the orbit and after replacing
$a$ by the orbital period $P$ we have $E = -\frac{1}{2} (2\pi G M_{\rm bh} / P)^{2/3}$
for $M_{\rm bh} \gg M_\ast$.
After one orbit the bound elements will return to $R_{\rm p}$ 
and efficiently loose their energy and angular momentum via stream-stream collision and 
finally be accreted to the black hole. The consequence is a luminous flare at $t_0$ 
with a peak in the UV/X-rays
and a spectrum which is characterized by a black-body temperature of a thick disk
or a spherical envelope at the tidal radius $R_{\rm t}$ from 
Eq.\,(\ref{Eqn:tidal_radius}):
\begin{equation}\label{Eqn:T_eff}
T_{\rm eff} 
            = \Bigg(\frac{L}{4 \pi \sigma R_\ast^2}\Bigg)^{1/4} \Bigg(\frac{M_{\rm bh}}{M_\ast}\Bigg)^{-1/6}.
\end{equation}

This scenario implicates two important conclusions for the light curve. 
First, if the energy distribution is uniform 
the mass return rate is determined by the relation between energy and period of the 
orbit $\mbox{d}M/\mbox{d}P \propto \mbox{d}E/\mbox{d}P \propto P^{-5/3}$. 
Assuming the time scale of the transformation of the orbital energy into radiation 
is short, i.e. the luminosity of the flare follows the accretion rate and the latter is given 
by the mass distribution of the return times, the flux density in the flare should be 
\begin{equation}
F \propto \Delta t^{-5/3},
\end{equation} 
where $\Delta t = t-t_{\rm d}$ is the time since the 
first passage of the pericentre.  
Numerical simulations have shown that this `standard' $\Delta t^{-5/3}$ light curve 
is a good approximation, at least for later stages (Evans \& Kochanek \cite{Evans89}). 
Close to the peak luminosity the light curve can be substantially shallower 
(Lodato et al. \cite{Lodato08}).
 
As another consequence, the time $\Delta t_0 = t_0 - t_{\rm d}$ 
the most-tightly bound material needs to fall back to
$R_{\rm p}$ is directly related to the mass of the black hole
\begin{equation}\label{Eqn:Delta_t_0}
\frac{\Delta t_0}{10^{-4}\,\mbox{yr}}  
\sim \beta^{-3} 
\Bigg(\frac{M_{\rm bh}\ R_\ast^3}{M_\ast^2\ k^3}\Bigg)^{1/2},\ \beta \equiv \frac{R_{\rm p}}{R_{\rm t}}
\end{equation}
with $M_{\rm bh}, M_\ast, R_\ast$ in solar units;   
$k$ depends on the spin-up state of the star with $k\sim 3$ 
for the likely case that the star is spun up to near break-up spin and
$k\sim 1$ if spin-up is negligible.

\begin{figure}[bhtp]   
\includegraphics{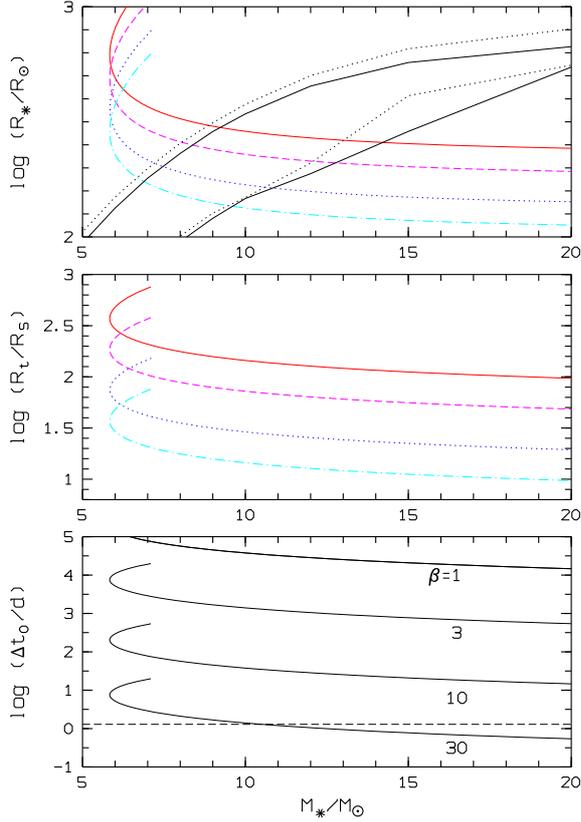}
\vspace{11.6cm}
\caption{
Stellar radius $R_\ast/R_\odot$ (top), ratio of tidal radius and gravitational radius 
$R_{\rm t}/R_{\rm S}$ (middle), and return time $\Delta t_0$ in days (bottom) as 
a function of the mass $M_\ast$ of the tidally disrupted star for 
$M_{\rm bh} = 1, 2, 5, 10\,10^8 M_\odot$ (top to bottom).
In the upper panel, the curves increasing from left to right are the 
$R_\ast$-$M_\ast$ relations for the stellar models from Salasnich et al. 
(\cite{Salasnich00}) for the base (bottom) and the top (top) of the red
giant branch, respectively, for solar metallicity (dotted: solar abundance ratios, 
solid: $\alpha$-enhanced abundances). 
The dashed horizontal line
in the bottom panel marks $\Delta t_0$ from the observed
light curve (Fig.\,\ref{fig:decline_slope}).
} 
\label{fig:TDE_model}
\end{figure}

In what follows, we shall check whether the total energy in the outburst of the Sharov\,21 
light curve, the decline of the outburst, and the time scales are consistent with the 
TDE scenario. The energy release in the flare is related to the mass of the star,
\begin{equation}
E_{\rm flare} = \eta f_{\rm acc} M_\ast c^2,
\end{equation} 
where $\eta$ is 
the efficiency of converting mass to radiated energy and $f_{\rm acc}$ is the fraction 
of mass of the star accreted to the black hole. For simplicity we assume 
$\eta f_{\rm acc} = 0.1$, which might be quite high 
(see Li et al. \cite{Li02}) but is not implausible, namely for 
$f_{\rm acc} \sim 0.5$ following Rees (\cite{Rees88}) and 
$\eta \sim 0.1 \ldots 0.4$ depending on the spin of the 
black hole. $E_{\rm flare}$ is obtained by integrating the
bolometric luminosity over the flare where $L_{\rm bol}(t)$ is computed from the 
monochromatic flux density, 
$F_{\nu,\rm flare}(t) = F_{\nu,\rm obs}(t) - \bar{F}_{\nu,\rm gs}$, with
$F_{\nu,\rm obs}(t)$ and $\bar{F}_{\nu,\rm gs}$ in the restframe. 
A black body spectrum is assumed with $T_{\rm eff}$ as a free parameter. 
The lowest possible stellar mass is $M_\ast \sim 6 M_\odot$ 
($E_{\rm flare} \sim 2\,10^{54}$ erg) corresponding to 
$T_{\rm eff} \sim 2\,10^4$ K where the black body spectrum peaks in the 
B band, i.e., at $\lambda = 4400$ \AA/$(1+z)$ restframe. 
For such a spectrum a colour index $B-R \sim 0.5$ mag 
is expected after foreground reddening, which is in line
with $B-R$ observed for Sharov\,21 in the flare (Fig.\,\ref{fig:colour_indices}). 

With $T_{\rm eff}, L_{\rm bol}$, and $M_\ast$ fixed, the stellar radius is 
constrained by Eq.\,(\ref{Eqn:T_eff}) for a given black hole mass $M_{\rm bh}$. 
In Fig.\,\ref{fig:TDE_model} (top), the resulting $(M_\ast,R_\ast)$ combinations
are compared with the stellar models from Salasnich et al. 
(\cite{Salasnich00}).
For $M_{\rm bh} \sim (1\ldots 10)\,10^8\,M_\odot$ we find
$M_\ast \sim (10\pm3)\,M_\odot$ and $R_\ast \sim (200\pm100)\,R_\odot$
which excludes main-sequence stars but not giants. For such giants, the tidal 
radius is clearly out of the gravitation radius, $R_{\rm S}$, 
of the black hole (Fig.\,\ref{fig:TDE_model}, middle), whereas main-sequence stars
of this mass range are again excluded as $R_{\rm t} < R_{\rm S}$.    
 
\begin{figure}[bhtp]   
\includegraphics{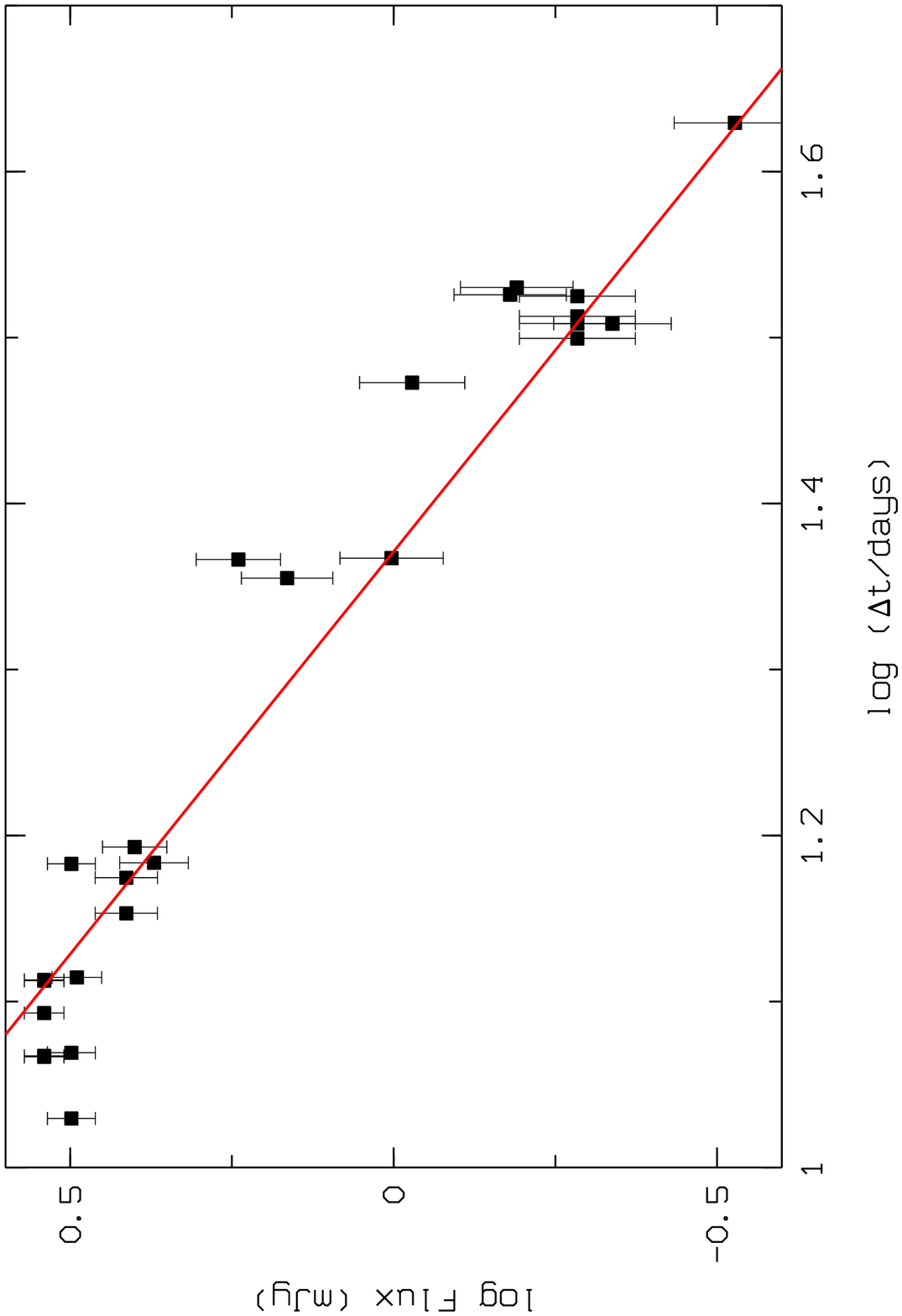}
\includegraphics{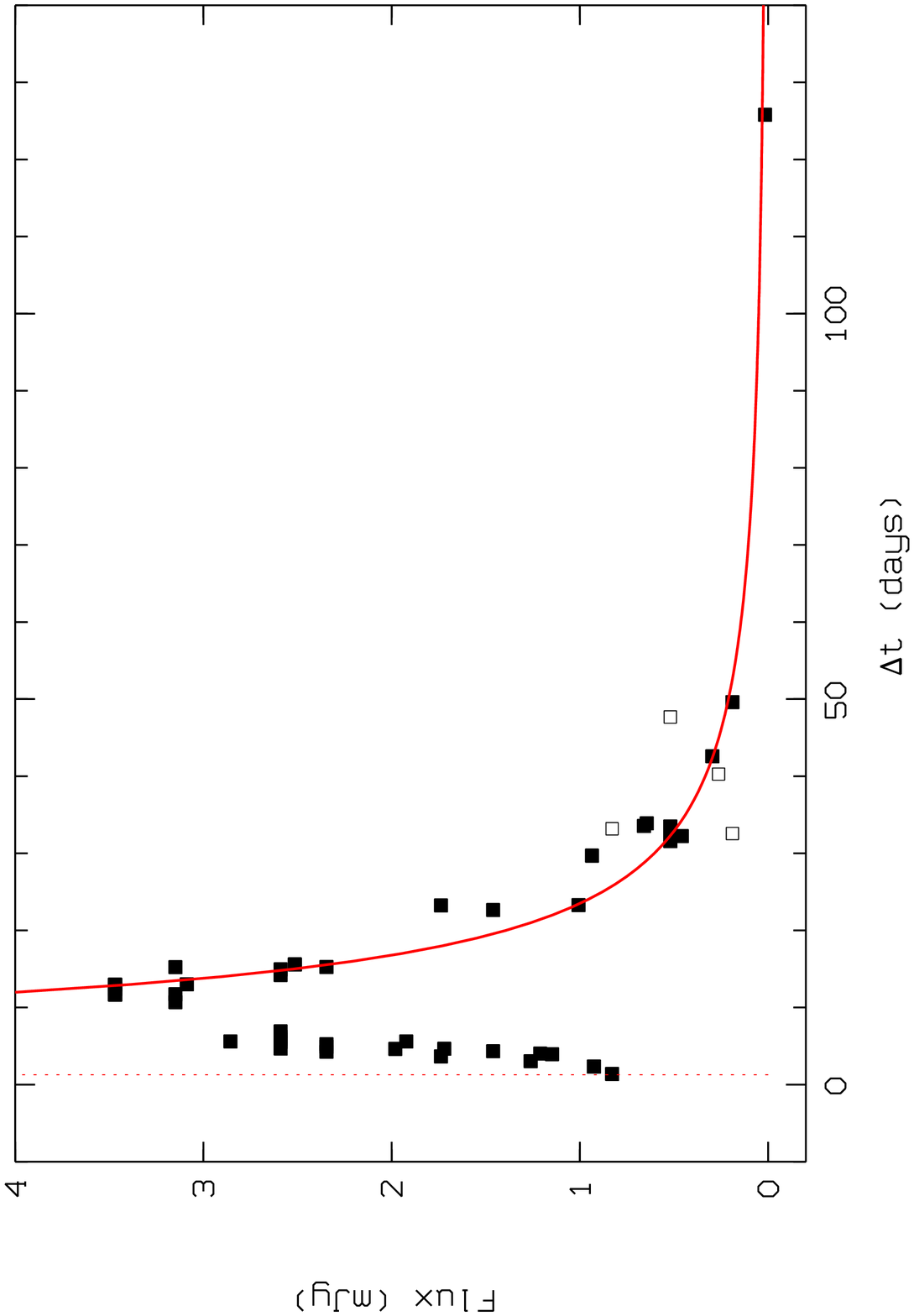}
\vspace{12.0cm}
\caption{
{\it Top:} Linear regression of the decline from the maximum in 
the double-logarithmic presentation of the light curve in the 
quasar restframe for the $\Delta t_0$ 
yielding a slope of $\gamma = -5/3$. 
{\it Bottom:} Fit of the light curve in linear presentation by the 
result from the regression in the top panel 
(open symbols: uncertain data; no error bars for lucidity). 
The vertical dotted line marks the onset of the flare at $t_0$.} 
\label{fig:decline_slope}
\end{figure}

In order to check whether the decline from the maximum follows the 
`standard' $\Delta t^{-5/3}$-law we perform linear regressions of
log\,$F_{\nu,\rm flare}$ as a function of log\,$\Delta t$
(Fig.\,\ref{fig:decline_slope}).
Whereas the onset of the flare at $t_0$ is defined by the
strong increase in the light curve at JD $\sim 2\,448\,897$, the beginning of the 
tidal disruption, $t_{\rm d}$, has to be considered as a free parameter. 
The resulting slope 
$\gamma = \mbox{d\,log} F_{\nu,\rm flare}/\mbox{d\,log} \Delta t$
depends on $\Delta\,t_0$ with  
$\gamma \sim -1.53 - 0.1 \Delta\,t_0$ (days).
The observed light curve is reasonably fitted by a power law with 
$\gamma = -5/3$ for $\Delta t_{0, \rm obs} \sim 1.3$ days and
is shallower at the beginning of the burst, as predicted by Lodato et al. 
(\cite{Lodato08}).

The bottom panel of Fig.\,\ref{fig:TDE_model} shows the fallback time $\Delta t_0$
as a function of $M_\ast$ for different values of the penetration factor $\beta$.
Note that $\Delta t_0$ from Eq.\,(\ref{Eqn:Delta_t_0}) 
is independent of $M_{\rm bh}$ because $R_\ast^3 \propto M_{\rm bh}^{-1}$ 
for given $(T_{\rm eff}, L, M_\ast)$ (Eq.\,\ref{Eqn:T_eff}). 
The short observed return time requires
$\beta \sim 30$ for $M_\ast = 10 M_\odot$.
With $R_{\rm p}/R_{\rm S} = (4.8, 2.4, 1.0, 0.5)$ 
for $M_{\rm bh} = (1, 2, 5, 10)\,10^8\,M_\odot$ 
and $M_\ast \sim 8\ldots 12\,M_\odot$,
the fundamental condition $R_{\rm p} > R_{\rm S}$ is matched for 
$M_{\rm bh} < 5\,10^8\,M_\odot$. 
Guillochon et al. (\cite{Guillochon09}) present highly resolved three-dimensional 
simulations of a tidally disrupted $1\,M_\odot$ 
solar-type star approaching a $10^6\,M_\odot$ black hole with $\beta = 7$, i.e. 
$R_{\rm p}/R_{\rm S} \sim 3$. Their results deviate from the case $\beta =1$
used in previous simulations. In particular, the simulated light curve shows 
a phase of smooth increase before the abrupt rise to the maximum, which is 
qualitatively in agreement with the behaviour of Sharov\,21 
(Fig.\,\ref{fig:outburst}).

The TDE scenario for the flare of Sharov\,21 is supported by
(1) total energy in the outburst,
(2) the shape of the light curve, and
(3) the time scale of the event.  
However, it must be noticed that the standard TDE theory refers 
to inactive black holes. The presence of a massive, self-gravitating accretion 
disk is expected to have a significant effect on the dynamics of the stellar 
tidal debris. Moreover, the description of the stellar orbit
for an encounter with the pericentre at only a few $R_{\rm S}$ 
requires a fully general-relativistic treatment which is, however, beyond the 
scope of this paper.

One may argue that the probability for the
capture of a $\sim 10 M_\odot$ giant star by a supermassive black hole 
is very low. Though this is certainly true, we emphasize that Sharov\,21-like quasar
flares are obviously extremely rare. Moreover it is worth mentioning  
(1) the high star formation rates and 
(2) the high fraction of massive stars in quasar host galaxies. 
It has been suggested for a long time that a luminous AGN phase is 
accompanied with or follows after an intense starburst, 
especially when both kinds of activity are induced by a galaxy merger 
(e.g., Sanders et al. \cite{Sanders88};
Granato et al. \cite{Granato04}; 
Springel et al. \cite{Springel05}).  
Lutz et al. (\cite{Lutz08}) derived strong evidence of the presence of intense star
formation (up to $\sim 3\,000\,M_\odot$ yr$^{-1}$) from {\it Spitzer} PAH
detections in 12 type\,1 quasar host galaxies at $z\sim 2$. Also from {\it Spitzer} IRS 
observations, Shi et al. (\cite{Shi09}) derived star formation rates on the 
level of luminous infrared galaxies ($10 \ldots 100\,M_\odot$ yr$^{-1}$)
for a complete sample of 57 type-1 quasar hosts at $z \sim 1$.
This high star formation activity appears to be concentrated in the circumnuclear 
region. In nearby Seyfert galaxies nuclear starbursts have been observed within $< 100$\,pc 
from the AGN (e.g., Davies et al. \cite{Davies07}; Watabe et al. 
\cite{Watabe08}). It is likely that more luminous AGNs are accompanied by more
luminous starbursts (Kawakatu \& Wada \cite{Kawakatu08}). 
The `radius of influence' of the black hole is 
$R_{\rm i} = G M_{\rm bh}/\sigma_\ast^2$, where $\sigma_\ast$ is the velocity dispersion 
of the surrounding stellar population. The $M_{\rm bh}$-$\sigma_\ast$ relation from 
Merritt \& Ferrarese (\cite{Merritt01}) yields 
$R_{\rm in} \sim 12 (M_{\rm bh}/M_\odot)^{0.58} 
\sim 30$\,pc for $M_{\rm bh} \sim 5\,10^8\,M_\odot$.
The typical distance a $10\,M_\odot$ giant can travel with a velocity $v = \sigma_\ast$ 
during its lifetime is $\sim 1$ kpc. 
(Note that the $M_{\rm bh}$-$\sigma_\ast$ relationship is not a strong 
function of redshift up to $z \sim 3$; Shields et al. \cite{Shields03}.)

Moreover, the central region of a galaxy provides a peculiar environment 
for star formation as is indicated by the 
high number of supergiants in our Galactic centre (e.g., 
Krabbe et al. \cite{Krabbe91}; 
Najarro et al. \cite{Najarro94};
Martins et al. \cite{Martins07}; 
Mauerhahn et al. \cite{Mauerhahn07})
with nearly a hundred massive stars within the central parsec.
Strong tidal forces, mass segregation, and other peculiar conditions may 
result in a relatively high
Jeans mass for collapsing cloud cores (Morris \& Serabyn \cite{Morris96}) 
and in a top-heavy initial mass function
(Figer et al. \cite{Figer02}; 
Paumard et al. \cite{Paumard06}; 
Bartko et al. \cite{Bartko09}). 
  
Gopal-Krishna et al. (\cite{Gopal08}) argue that the gas from TDEs may be a major 
factor for the abortion of radio jets in quasars. If the radio loudness is 
related to $M_{\rm bh}$, as seems to be indicated by observations, the 
radio-loudness dichotomy can be explained by the existence of a critical 
black hole mass for the tidal disruption of solar type stars. As noted by these 
authors, the tidal disruption of giant stars is then required to explain the 
association of a few radio-quiet quasars with black holes of high masses.

\subsection{Other interpretations}\label{other_interpretations}

In the pioneering work of Cannon et al. (\cite{Cannon68}, \cite{Cannon71}),
the class of quasars with variability amplitudes of the order of one magnitude and 
with often rapid fluctuations were called optically violently variables (OVVs). 
OVV AGNs are all radio-loud and their strong variability is believed
to be due to relativistic beaming (Gaskell \& Klimek \cite{Gaskell03}, and
references therein). According to the current zoology of AGN types, OVVs 
constitute a subtype of blazars, i.e. of jet-dominated active galaxies viewed 
close ($\sim 15\degr$) to the axis of a relativistic jet 
(Urry \& Padovani \cite{Urry95}).  Viewing the jet almost directly head-on 
results in a great magnification of variations in the flux from the jet.
Blazars vary dramatically on a wide range of time scales from hours to years.
In this context, the colour dependence of the variability found for the two 
strongest variable OVV quasars from the SDSS S82 (Sect.\,\ref{Rareness}) can 
be easily understood assuming that the spectral energy distribution is dominated 
by the synchrotron peak in the radio and infrared. This is the case for 
low-peaked BL Lac objects and is typical also for flat-spectrum radio quasars:  
When the synchrotron emission from the jet makes a significant contribution 
which increases with wavelength, 
the optical spectrum becomes redder when the jet is brighter and the variability
is stronger at longer wavelengths. 
Here we argue that Sharov\,21 is not an OVV quasar in the `classical' sense:
(1) The strong flux variation is limited to one single event, 
there is no evidence of strong variability on a wide range of time scales. 
(2) The spectral behaviour of the variability is typical of 
radio-quiet quasars (Sect.\,\ref{Rareness}).
(3) Sharov\,21 is not radio-loud.

Variability of non-blazar AGNs has been attributed to various processes like 
oscillations in the accretion flow (Igumenshchev \& Abramovicz \cite{Igumenshchev99}) 
or in the jet (Hughes et al. \cite{Hughes98}). Both models
produce quasi-periodic features in the light curve where the maxima 
have approximately the same widths as the intervals in between, which is
contrary to what we see in Sharov\,21. Avalanche flows in the accretion disk 
provide another explanation (Tekeuchi et al. \cite{Takeuchi95}).
The profile of a single flare in simulated light curves 
(Kawaguchi et al. \cite{Kawaguchi98}) is characterized 
by a gradual increase in brightness followed by a sudden decline. It is not clear 
whether the rapid increase to the maximum can be explained by such a model.

An alternative explanation of AGN variability is based on the superposition of
uncorrelated events such as supernova explosions (Terlevich \cite{Terlevich92}),
star collisions (Courvoisier et al. \cite{Courvoisier96}; 
Torricelli-Ciamponi et al. \cite{Torricelli00}), or colliding clumps in a 
clumpy accretion flow (Courvoisier \& T\"urler \cite{Courvoisier05}).
First, we compare the energy in the flare of Sharov\,21 with the radiated energy
of supernovae. SN\,2006gy has been considered the most luminous supernova with a
total radiated energy $E_{\rm SN, rad} \sim 10^{51}$\,erg 
(Smith et al. \cite{Smith07}; Ofek et al. \cite{Ofek07}).
For the flare of Sharov\,21 we estimate $E_{\rm flare} \sim 10^{54\ldots55}$\,erg,
dependent on the spectral energy distribution, hence 
$E_{\rm flare} \sim 10^{\,3\ldots4} E_{\rm SN, rad, max}$. 
It was the aim of the collision scenario to explain both the ultraviolet 
emission and its variability. As a modified version, such events could be 
rare and contribute flares to a continuum source the emission and variability 
of which are dominated by other processes.
In fact, the star collision scenario is distantly related to 
the TDE model. The maximum amplitudes of the flares in the light curves 
presented by Torricelli-Ciamponi et al. are much too low compared to
Sharov\,21. However it remains to be tested if the model parameters
can be adjusted to fit our light curve.  

In another class of models strong flares are explained by processes in binary black 
holes. Lehto \& Valtonen (\cite{Lehto96}) suggest a model where they 
associate the quasi-periodic outbursts of the OVV quasar OJ\,287 to the times when
the secondary crosses the accretion disk of the primary black hole. When 
this happens the secondary pierces a channel through the disk, the gas is heated 
in the channel, flows out, and radiates strongly. The model was demonstrated to be 
successful for OJ\,287 (Valtonen et al. \cite{Valtonen08}). Other models attribute 
radiative outbursts to the tidal perturbation of the accretion flow due to 
the secondary (Sillanp\"a\"a et al. \cite{Sillanpaa88}), or the   
sweeping of a processing relativistic beam across the line of sight, in analogy
to SS\,433 (Katz \cite{Katz97}; Liu \& Chen \cite{Liu07}).
OJ\,287 is the best-studied quasar with a quasi-periodicity in the light curve.  
Its outbursts come in 12\,yr intervals corresponding to 9\,yr in the restframe. 
Our light curve covers, in the restframe, only $\sim 10$ yr 
before the outburst and $\sim 5$ yr after. 
If orbital time scales longer than about one decade are 
not atypical of supermassive black hole
binaries, it cannot be excluded that the flare of Sharov\,21 is related to such a 
process.  It would be important to 
find more historical observations in plate archives which yield
an extension of the light curve further back. This however is much more difficult than 
for OJ\,287 which is $\sim 5$ mag brighter than Sharov\,21.   

Finally, we briefly mention the possibility that the quasar
was by chance superposed by a nova in M\,31. This hypothesis was tested already
by Sharov et al. (\cite{Sharov98}). From the measurement of the position 
before, during, and after the flare they find that the derived coordinates
are in excellent agreement with each other and conclude that 
the flaring object is ``coincident with the star that was visible at its place 
in 1982-1997''.

%
%
\section{Summary and conclusions}\label{conclusion}
%
%

Sharov\,21, originally discovered by Nedialkov et al. (\cite{Nedialkov96}),
was previously classified as a remarkable nova in M\,31 (Sharov et al. 
\cite{Sharov98}). Here we have shown that it is a
typical type\,1 quasar at $z = 2.109$ 
seen through the disk of M\,31 and showing an exceptionally strong UV flare. 
We created a significantly improved long-term light curve based on 
archival data, data from the literature, and targetted new observations. 
Compared to the original data given by Sharov et al., 
the new light curve has a $\sim 20$ yr longer baseline and a better sampling. 
Altogether, more than $10^3$ single exposures from 15 wide field telescopes 
are included resulting 
in detections at 221 epochs from 1948 to 2009, with a relatively good 
time coverage after 1961. The data material is completed by a large 
number of observations without detection of the quasar but with useful upper 
brightness limits for the time interval from 1900 to 2009. 
Based on this data, we subdivide the light curve in two phases: 
a rather quiet ground state with $\bar{B} \sim 20.5$ 
for at least 98\% of the time and a strong outburst in 1992 with an increase 
of the UV flux by a factor $\sim 20$.The variability in the ground state does 
not significantly differ from that of other radio-quiet quasars of comparable 
redshift and luminosity. A black hole mass of $5\,10^8\,M_\odot$ is estimated 
from the \ion{C}{iv} line, 
corresponding to an Eddington 
ratio of $\sim 0.6$ for the ground state. 
By the comparison with $\sim 8\,000$ quasars in the stripe 82 of the 
SDSS on a 7\,yr-baseline (Bramich et al. \cite{Bramich08})
and more than 300 VPMS quasars  (Meusinger et al. \cite{Meusinger02}, 
\cite{Meusinger03}) with light curves having a time-baseline comparable to Sharov\,21
we have demonstrated that the strong UV flare of Sharov\,21 
is very unusual for radio-quiet high-redshift quasars. 
We conclude that such a rare feature 
is the result of a rare event. As such 
we suggest two scenarios:  (1) gravitational microlensing due 
to a star in M\,31 and (2) a tidal disruption event (TDE) of a star close 
to the supermassive black hole of the quasar.  

In the TDE scenario, the total energy in the outburst can be explained by 
the disruption of a $\sim 10\,M_\odot$ giant star if we `optimistically' 
assume that half of the disrupted star is accreted to the
black hole and $\sim 20$\% of the accreted energy can be radiated away.
The flare profile shows a sudden increase to the maximum followed
by a decline which is reasonably fitted by the $\Delta t^{-5/3}$ power law
predicted by the standard TDE model. The short time span between the 
beginning of the tidal disruption and the beginning of the flare,
$\Delta t_0 \sim 1.3$\,days, requires an ultra-close encounter 
with $\beta \sim 30$ corresponding to a stellar orbit with the pericentre 
at only a few $R_{\rm S}$.  The present study has not taken 
general-relativistic effects into account. Moreover, it is unclear how the 
TDE scenario is affected by the presence of a massive accretion disk. 
Microlensing by a star in M\,31 is a plausible alternative explanation. 
Though the detection of a microlensed quasar behind M\,31 over
half a century is not unlikely, high-amplification events 
corresponding to the flare of Sharov\,21 are very rare.
We apply the point source-point lens approximation to model the light curve.
Assuming a transverse velocity of 300\,km\,s$^{-1}$, an acceptable fit
is achieved for a low-mass binary with $0.3\,M_\odot$ and $0.1\,M_\odot$.
The observed light curve of Sharov\,21 is roughly fitted by either of the two 
scenarios, but more detailed modelling is necessary to decide if the flare 
can be reproduced accurately. 
Finally, we cannot exclude that the flare is part of a quasi-periodic 
activity similar to OJ\,287 (Valtonen et al. \cite{Valtonen08}) on an intrinsic
time scale of $\ga10$ yr. 
The remarkable quasar Sharov\,21 obviously merits further efforts, both for the 
completion of the light curve and for its modelling. 

Finally, we notice that there is an interesting application of 
Sharov\,21-like flares. Projects like PTF, Pan-STARRS, and LSST 
will probably discover several such events. Due to their enormous 
luminosity and long time scale in the observer frame, such flares can provide 
background light sources for intervening matter and create thus 
interesting opportunities for high-resolution spectroscopy of matter at 
large distances. This argument has been presented by Quimby et al. 
(\cite{Quimby09}) in the context of the brightest supernovae from the PTF
but holds even more for quasar-flares comparable to that one of
Sharov\,21.

\begin{acknowledgements}
The anonymous referee is greatly acknowledged for his constructive 
criticism which helped to improve the paper.
We thank S. Komossa for helpful comments on an earlier draft version 
of this paper and A. Kann, S. Klose, M. R\"oder, S. Schulze, and B. Stecklum for 
contributing observations with the Tautenburg Schmidt CCD camera.
We are grateful to M. Demleitner, Astronomisches Rechen-Institut,
Heidelberg, for installing and managing the HDAP data base as part
of GAVO, to G. Langer and L. Siegwald, LSW HD, for scanning the
Heidelberg and Calar Alto plates, and to O. Stahl, LSW HD, for manifold
support, particularly with data handling. The HDAP project is financed
by the Klaus Tschira Foundation, Heidelberg under contract
no. 00.071.2005. The Tautenburg plate scanner was financially supported 
by the Deutsche Forschungsgesellschaft under grants Me1350/3 and Me1350/8.
M.H. acknowledges support from the BMWI/DLR, FKZ 50 OR 0405.
This research  draws upon data provided by Philip Massey as distributed by the 
NOAO Science Archive. NOAO is operated by the Association of Universities 
for Research in Astronomy (AURA), Inc. under a cooperative agreement with the 
National Science Foundation.
We make also use of data obtained from the Isaac Newton Group Archive which 
is maintained as part of the CASU Astronomical Data Centre at the Institute of 
Astronomy, Cambridge, and used the facilities of the Canadian Astronomy Data 
Centre operated by the National Research Council of Canada with the support 
of the Canadian Space Agency.
Some of the data presented in this paper were obtained from the Multimission 
Archive at the Space Telescope Science Institute (MAST). STScI is operated by 
the Association of Universities for Research in Astronomy, Inc., under NASA 
contract NAS5-26555. Support for MAST for non-HST data is provided by the NASA 
Office of Space Science via grant NAG5-7584 and by other grants and contracts.

\end{acknowledgements}


{}

\end{document}